\documentclass[pdflatex,sn-basic,iicol]{sn-jnl}

\usepackage{subcaption}

\jyear{2021}%

\theoremstyle{thmstyleone}%
\theoremstyle{thmstyletwo}%

\theoremstyle{thmstylethree}%

\begin{document}

\title[Article Title]{Skydiving Technique Analysis from a Control Engineering Perspective: Developing a Tool for Aiding Motor Learning}


\author*[1]{\fnm{Anna} \sur{Clarke}}\email{anna\_shmaglit@yahoo.com}

\author[2]{\fnm{Per-Olof} \sur{Gutman}}\email{peo@technion.ac.il}


\affil*[1]{\orgdiv{Technion Autonomous Systems Program (TASP)}, \orgname{Technion - Israel Institute of Technology}, \orgaddress{ \city{Haifa}, \postcode{32000}, \country{Israel}}}

\affil[2]{\orgdiv{Faculty of Civil and Environmental Engineering}, \orgname{Technion - Israel Institute of Technology}, \orgaddress{ \city{Haifa}, \postcode{32000}, \country{Israel}}}


\abstract{This study offers an interdisciplinary approach to movement technique analysis, designed to deal with intensive interaction between an environment and a trainee. The free-fall stage of skydiving is investigated, when aerial manoeuvres are performed by changing the body posture and thus deflecting the surrounding airflow. The natural learning process of body flight is hard and protracted since the required movements are not similar to our daily movement repertoire, and often counter-intuitive. The proposed method can provide a valuable insight into the subject’s learning process and may be used by coaches to identify potentially successful technique changes. The main novelty is that instead of comparing directly the trainee’s movements to a template or a movement pattern, extracted from a top-rated athlete in the field, we offer an independent way of technique analysis. We incorporate design tools of automatic control theory to link the trainee’s movement patterns to specific performance characteristics. This makes it possible to suggest technique  modifications that provide the desired performance improvement, taking into account the individual body parameters and constraints. Representing the performed manoeuvres in terms of a dynamic response of closed-loop control system offers an unconventional insight into the motor equivalence problem.
The method is demonstrated on three case studies of aerial rotation of skilled, less-skilled, and elite skydivers.}




\maketitle

\section{Introduction}\label{sec1}

Skydiving involves skilful body movement to control free-fall. While skydivers fall near terminal vertical velocity they alter their body posture, thus deflecting the air flow and initiating a great variety of solo and group aerial manoeuvres. Mastering body flight is naturally difficult and protracted. The reason is that the required body movements are not similar to our daily movement repertoire, and are often counter-intuitive. The stressful environment, causing muscle tension and blocking kinaesthetic feedback, exteroceptive sensory overload, and a very limited prior information on desired body postures constitute additional learning challenges. Skydiving body movements depend highly on individual anthropometric, clothing, and equipment factors. Therefore, coaches cannot predict what movements will be required from trainees for performing even the simplest manoeuvres, like falling straight down. Moreover, they cannot describe even their own joint movements, since their bodies perform them automatically before being aware that posture adjustments are required. These joint rotations are usually small, often unnoticeable on a video recording. The coaches give only general recommendations regarding relaxation, visualization, and focus of attention. For a qualitative description of common skydiving moves, see \cite{bodyflight}. In \cite{article} a simulation of free-fall dynamics was developed. It predicts the skydiver's motion in the 3D space given a sequence of body postures, and, therefore, can be utilized for studying the skydiving technique.

Technique is the way an athlete's body executes a specific sequence of movements \cite{gloersen2018technique, lees2002technique}. It can be viewed as an athlete’s repertoire of movement patterns: combinations of body Degrees of Freedom (DOFs) that are activated synchronously and proportionally, as a single unit. From the perspective of dynamical systems theory, motor learning is the process during which these movement patterns emerge \cite{sport_schmidt2008motor}. First, the patterns are simple (coarse), providing just the basic functionality. As learning continues, the movement patterns become more complex (fine), providing adaptation to perturbations and uncertainties, and improved performance. We offer a new method to link the acquired behaviours of the dynamical system to the specific changes in the movement patterns, and predict how the movement patterns can be altered to provide desired performance. Existing quantitative methods of technique analysis attempt to provide similar insight. For example, deterministic modelling is widely used for analysing swimming, athletics, and gymnastics \cite{chow2011use}. It determines the relation between the performance measure and mechanical variables, e.g. horizontal velocity at take-off point during a somersault. It thus identifies parameters important for performance and their target values. Some of those parameters may be related to joints positions and rotation speeds, but the full body motion sequence is usually not provided.  Additionally, sometimes multiple techniques can be adopted to generate some of the mechanical variables due to human motor equivalence. 

Other conventional technique analysis methods, such as variability/ temporal analysis and phase portraits, suffer from some pitfalls: collection of only partial biomechanical data that are insufficient for analysis of the whole skill, and focusing only on some variables or aspects believed to most affect performance or best represent learning progress. See e.g.: \cite{davids2005applications, federolf2012holistic, hamill2000issues, lamb2014use, muller2004decomposition, myklebust2016quantification, scholz2014use, stergiou2016nonlinear, sternad2018s}.

Most often, the methods analysing sports techniques focus on comparing the movement of the trainee to a template/ coach/ top rated athlete (e.g. \cite{ghasemzadeh2011coordination, federolf2014application, gloersen2018technique}). Moreover, required movements are usually well known and have large amplitude, as in kicks, swings, poles placement. Thus, the analysis focuses mainly on the timing of those movements and their small variations (\cite{holmberg2005biomechanical}). 

Our method aims to be wider-ranging. It is designed to analyse mathematically, without templates, the technique of a previously unexplored activity that is highly dependent on individual body properties, taking into account that all body parts are involved and slight posture variations might have a large influence. Trainees might have different physical condition, age, and even disabilities. As an illustration, skydiving attracts very diverse participants. The reason is that free-fall manoeuvres do not require a significant muscle power or physical fitness. All manoeuvres are generated by aerodynamic forces and moments when a skydiver deflects the air flow around his body by changing the posture. Consequently, our method is designed to deal with an intensive interaction between environment and trainee. The method has one sports-specific part (modelling), and one general part in which the predicted dynamic responses are used for adjusting the technique to make it more efficient. The model computing these responses is driven by the recorded biomechanical measurements. Thus, we adopt an interdisciplinary approach, integrating dominant motor learning concepts with the analytical tools of automatic control theory. We hypothesise that an individual skydiving technique can be assessed if it is interpreted as an actuation strategy of a closed loop feedback control system comprising the trainee's body and the environment.

\section{Materials and Methods}\label{sec2}

One professional skydiving instructor with over 30 years of experience and 15000 jumps, one amateur skydiver with less than 10 years of experience and 1000 jumps, and one elite skydiver competing in the discipline Relative Work (RW) with 25 years of experience, 2500 jumps and numerous hours of wind tunnel training participated in the study. The three participants will be hereafter referred to as: Instructor, Student, and Elite Skydiver, respectively. The Student is still pursuing improvement of his personal technique. Since each jump has about 30 seconds of training time, his total free-fall manoeuvring experience does not exceed 8.5 hours. Therefore, his movement patterns are coarse, and some of the body DOFs are still locked. 
The flow of the study included collecting the body movement data from the participants, processing it by the means of Principal Component Analysis (PCA), and analysing the results by the means of a Skydiver Simulator developed for this purpose.  

\subsection{Dynamic simulation of the human body in free-fall}\label{subsec20}

The dynamic simulation of the human body in free-fall is essential for our technique analysis method. Therefore, a simulation that receives a sequence of body postures and computes position, orientation, and linear and angular velocities of the skydiver model in a 3D world was developed, see \cite{article}. The inputted postures can be recorded, transmitted in real-time, as well as synthetically generated, or even inputted via a keyboard using a graphical user interface, specifically designed for this purpose. The simulation is implemented in Matlab and it has a continuous graphical output, which shows a figure of a skydiver in its current pose moving through the sky. The sky has a grid of equally spaced half-transparent dots, so that the skydiver’s manoeuvres can be easily perceived by the viewer. The modules comprising our skydiving simulator are briefly described below, while the exact equations can be found in our previous work (\cite{article}).

The \textit{biomechanical model} represents the body segments in terms of simple geometrical shapes and calculates the local centre of gravity and principal moments of inertia for each segment. A set of rotation quaternions linking each two segments enables computation of the overall centre of gravity, inertia tensor, and their time derivatives. The model has to be provided with a set of parameters, expressing body size, shape, and weight of the skydiver under investigation. The \textit{dynamic equations of motion}, derived following the Newton-Euler method, provide six equations: 3D forces and moments. The \textit{kinematic model} computes the body inertial orientation, and angles of attack, sideslip, and roll of each segment relative to the airflow. These angles are used in the \textit{aerodynamic model} to compute drag forces and aerodynamic moments acting on each segment. The total aerodynamic force and moment together with the gravity forces are substituted into the equations of motion. The aerodynamic model is formulated as a sum of forces and moments acting on each individual segment, modelled similar to aircraft aerodynamics - proportional to velocity squared and to the area exposed to the airflow. The model includes six aerodynamic coefficients that were experimentally estimated, and has to be provided with a set of configuration parameters specific to the skydiver under investigation (type of parachute, helmet, jumpsuit, and weight belt).

The skydiving simulator output was experimentally verified: Various skydiving manoeuvres in a belly-to-earth pose were performed by different skydivers in a wind tunnel and in free-fall, and the recorded posture sequences were fed into the simulator. The six tuning parameters related to the aerodynamic model (maximum lift, drag, and moment coefficients; roll, pitch, and yaw damping moment coefficients) were selected so that all the manoeuvres were closely reconstructed. The errors RMS in angular and linear (horizontal and vertical) velocities were 0.15 rad/s, 0.45 m/s (horizontal), and 1.5 m/s (vertical), while the velocities amplitudes were 7 rad/s, 15 m/s, and 65 m/s, respectively.

\subsection{Equipment}\label{subsec21}

The X-Sens body movement tracking system \cite{sensor_roetenberg2009xsens} was chosen for the purpose of getting an accurate measurement of a full body posture, meaning knowing the relative orientation (three DOFs) of all relevant body segments. It provides a suit with 16 miniature inertial sensors that are fixed at strategic locations on the body. Each unit includes a 3D accelerometer, 3D rate gyroscope, 3D magnetometer, and a barometer. The X-Sens motion tracking suit can be worn underneath conventional skydiving gear. It has a battery and a small computer located on the back and not restricting the skydiving-specific movements. The X-Sens output is transmitted or recorded at 240 Hz. Each measurement set includes the orientation of 23 body segments (pelvis, four spine segments, neck, head, shoulders, upper arms, forearms, hands, upper legs, lower legs, feet, toes) relative to the inertial frame, expressed by quaternions. The measurements accuracy is less than 5 degrees RMS of the dominant joint angles \cite{schepers2018xsens}.

 The experiments took place in a wind tunnel (diameter 4.3 m, height 14 m), which is a widely used skydiving simulator, where the air is blown upwards at around 60 m/s - an average terminal vertical velocity of skydivers in a belly-to-earth pose. The human body floats inside the tunnel replicating the physics of body flight. The experiments were videoed for future reference, see Online Resources \ref{res1}-\ref{res3}. 

\subsection{Measurement procedure}\label{subsec22}

The Instructor, Student, and Elite Skydiver were in turn equipped with the X-Sens suit and performed the calibration procedures defined by X-Sens that are necessary for the convergence of the measurement system. The participants' body segments and height were measured and inputted into X-Sens software, as well as into the Skydiving Simulator, described Sect. \ref{subsec20}, along with weight, helmet size, and jumpsuit type, worn on top of the X-Sens suit.

Next, the participants entered the wind tunnel and performed 360 degrees turns to the left and to the right in a belly-to-earth pose during two minutes – a typical wind tunnel session.  The participants were instructed to perform as many turns as they could within this session, while preventing any horizontal or vertical displacement relative to the initial body position in the centre of the tunnel. This is an essential skydiving skill.

\subsection{Data processing}\label{subsec23}
\subsubsection{Extracting and visualising the movement patterns}\label{subsubsec231}

\begin{figure}[h]%
\centering
\includegraphics[width=0.4\textwidth]{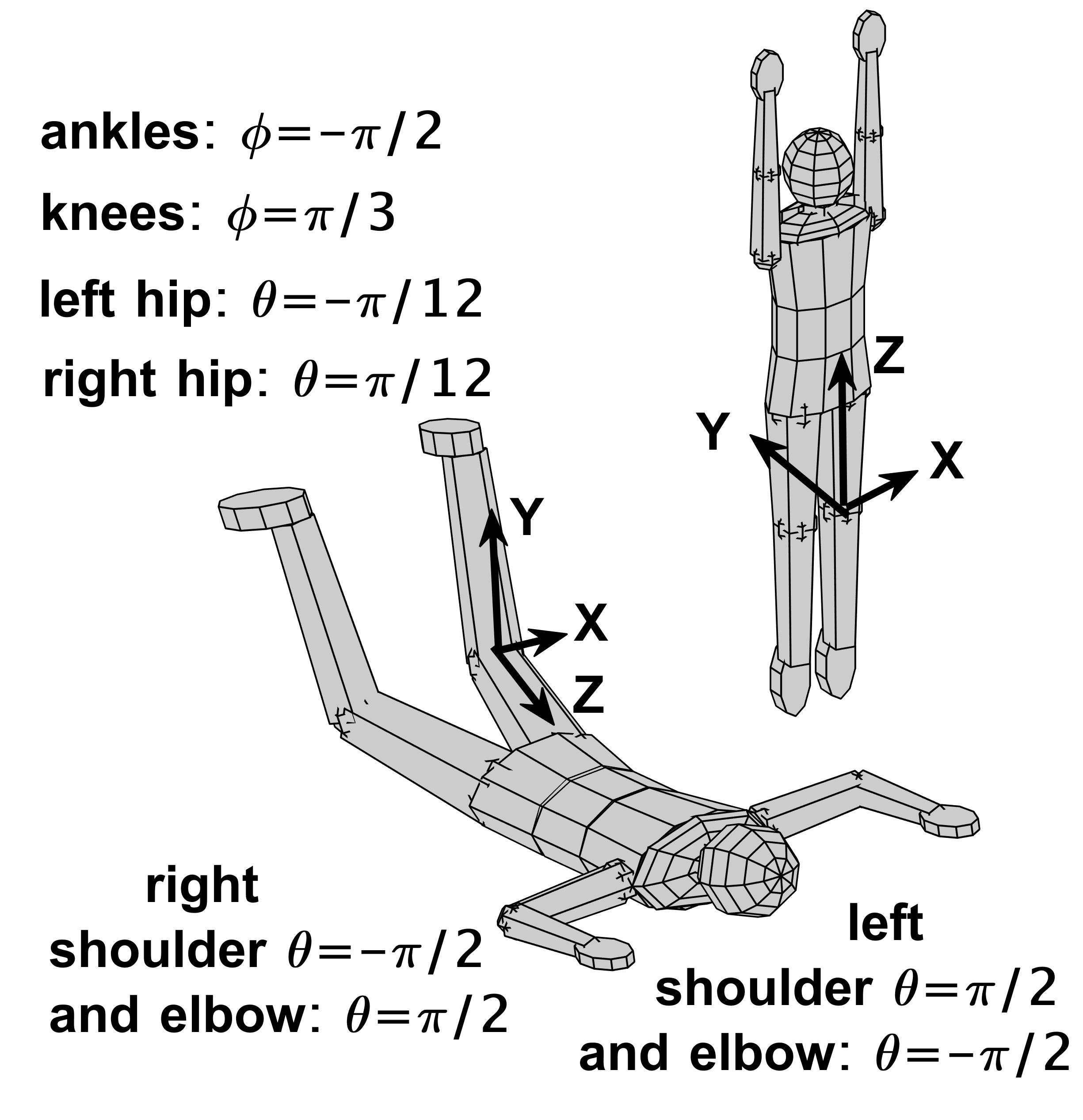}
\caption{Pose defined by all zero Euler angles (top), and a standard neutral pose (bottom). For each limb the Euler angles $\psi$, $\theta$, $\phi$ are rotations around $Z$, $Y$, $X$ axis of the coordinate system attached to the parent limb}\label{fig1}
\end{figure}

The movement patterns were extracted from the X-Sens data by means of PCA - a method widely used for analysing human motion in general and sports technique in particular, see \cite{federolf2014application} for a review. PCA uses Singular Value Decomposition (SVD) of the recorded motion data in order to find movement patterns. In our case, X-Sens data consists of inertial orientation of 23 body segments. For skydiving in a belly-to-earth pose such a detailed posture measurement is superfluous, hence some segments can be united. Our skydiver body configuration \cite{article} is defined by 16 rigid segments (pelvis, abdomen, thorax, head, upper arms, forearms, hands, upper legs, lower legs, and feet), and 15 joints (lumbar, thorax, neck, shoulders, elbows, wrists, hips, knees, ankles). The relative orientation of connected segments is defined by three Euler angles (the sequence of rotation is shown in Fig. \ref{fig1}), which are computed from the recorded X-Sens quaternions. 
The data matrix is constructed from 3$\cdot$15 rows and 120$\cdot$240 columns, since there are 15 joints, each with 3 DOFs, the sampling frequency is 240 Hz, and each tunnel experiment is 120 s. From each data row we subtract its mean, whereas normalization of rows is not required in our case, since all rows have the same units and the differences in variability are meaningful. Computing the SVD results in a diagonal matrix $S$ and unitary matrices $U$ and $PC$, such that: 
\begin{equation}
    data^T=U\cdot S\cdot PC^T
     \label{eq1}
\end{equation}
The matrix $PC$ of size 45 by 45 contains the Principal Components: eigenvectors that define the components of the skydiver’s turning technique, hereafter referred to as \textit{movement components}. The 45 diagonal elements of $S$ are the eigenvalues corresponding to each movement component, with more dominant components having larger eigenvalues. Projecting the original data into the principal components provides the time evolution coefficients of each of the components, i.e. their engagement as a function of time. These coefficients have a role of \textit{control signals} in Automatic Control Systems, and are the rows of the following matrix: 
\begin{equation}
    ControlSignals=PC^T\cdot data
    \label{eq2}
\end{equation}
The skydiver’s pose at time $t$ is composed as:
\begin{equation}
    \begin{gathered}
    pose(t)=N_{pose} + \sum_{i=1}^{45}{\alpha_i(t)\cdot PC_i} \\
    \alpha_i(t)=ControlSignal_i(t) \\
     N_{pose}[j] = \frac{1}{120 \cdot 240}\sum_{k=1}^{120 \cdot 240}{data[j,k]} \\
     \text{for  } j \in [1, 45]
    \end{gathered}
     \label{eq3}
\end{equation}
where $ControlSignal_i(t)$ is value from row $i$ of matrix $ControlSignals$ corresponding to time $t$, i.e. from column $t \cdot 240$; $N_{pose}[j]$ is the entry $j$ of the column vector $N_{pose}$ containing the mean of the sequence of recorded postures and, thus, representing the \textit{neutral pose}; and $PC_i$ is the eigenvector (with norm 1) defining the movement component $i$, i.e. column $i$ of matrix $PC$. Thus, the control signal $\alpha_i(t)$ is the angle in radians of the movement component being engaged, and will be referred to as the \textit{pattern angle}, since each movement component defines a certain pattern of coordination contributing to the total variance of the posture.

The Skydiving Simulator can be used to animate a specific movement component multiplied by any synthetically constructed or recorded control signal. We often use a sine wave with amplitude $A=1$ rad and frequency $f=1$ Hz to animate PCA movement components:
\begin{equation}
\begin{gathered}
    pose_i(t)=N_{pose} + A\cdot sin(\omega \cdot t)\cdot PC_i \\
    \omega = 2\cdot \pi \cdot f
    \end{gathered}
    \label{eq4}
\end{equation}
This makes it possible to give an athlete a very detailed visual feedback of coordinative structures comprising his movement repertoire. 

The Skydiving Simulator fed by the sum of all identified PCA movement components, multiplied by the corresponding control signals, will reconstruct the performed manoeuvres. However, if we choose to input specific PCA movement components, multiplied by the recorded or synthetically generated control signals, we can obtain answers to a great variety of interesting questions related to the skydiving skill and its acquisition process. Below we describe the analysis configurations that provided the most useful insight.

\subsubsection{Manoeuvre generated by a given PCA movement component}\label{subsubsec232}

Each movement component reconstructed by PCA has a corresponding normalized eigenvalue in the range $[0, \, 1]$. Only a few PCA components are expected to have eigenvalues greater than 0.2, hereafter referred to as \textit{dominant components}. An accepted interpretation of these values is the percentage of total posture variability a movement component is responsible for, see \cite{gloersen2014quantitative, hollands2004principal}. We seek a deeper insight: describing the exact role of a given PCA movement component in the overall manoeuvre. This is achieved by feeding the Skydiver Simulator by each dominant movement component  with:
\begin{enumerate}
    \item a synthetic periodic control signal (Eq. \eqref{eq4})
    \item a step input: $pose_i(t)=N_{pose} + A \cdot PC_i$
\end{enumerate}
Notice, that in the latter case the pose is constant in time and its difference from the neutral pose is $A \cdot PC_i$. For $A=1$ rad  this quantity will be referred to as a movement pattern with dimensions, as opposed to a dimensionless eigenvector $PC_i$. The frequency and amplitude of the control signal in Eq. \eqref{eq4} are chosen to obtain significant and yet feasible pose changes in time.

\subsubsection{PCA movement components required to reconstruct the original manoeuvre}\label{subsubsec233}
The following method provides an insight on how many PCA movement components are required to reconstruct the manoeuvre under investigation. It suggests feeding the skydiver simulator by a different amount of dominant PCA movement components with the corresponding control signals from the experiment:
\begin{equation}
    pose_n(t) = N_{pose} + \sum_{i=1}^{n}{\alpha_i(t)\cdot PC_i}
    \label{eq5}
\end{equation}
where $n$ is the number of chosen PCA movement components. For $n=45$ we obtain $pose_{45}(t)$ - an exact pose the skydiver had during the experiment at every instant of time. Since the manoeuvre in our experiment was turning, its outcome can be described by yaw rate. In order to compare the outcome produced by different pose inputs, we define the total discrepancy between the yaw rate profiles $\Omega_{45}(t)$ and $\Omega_{n}(t)$ resulting from $pose_{45}(t)$ and from $pose_{n}(t)$, respectively, accumulated during the time of the experiment $T=120$ s, which included $N=\frac{120}{\delta t}$ steps, $\delta t=\frac{1}{240}$  s:
\begin{equation}
     Err_n=\sum_{k=1}^{N}{\lvert \Omega_{45}(k)-\Omega_{n}(k) \rvert \cdot \delta t}
     \label{eq6}
 \end{equation}
The similarity of the outcome manoeuvres can be defined as: 
\begin{equation}
     \begin{gathered}
     Sim_n=1-{Err_n}/{Err_0} \\
     Err_0=\sum_{k=1}^{N}{ \lvert \Omega_{45}(k) \rvert \cdot \delta t}
     \end{gathered}
      \label{eq7}
 \end{equation}
 where $Err_0$ is the error if the skydiver doesn't manoeuvre at all by keeping the neutral pose at all times. Thus, if the component $PC_1$ had no influence on the yaw rate, $Sim_1$ would be zero, whereas, if only $PC_1$ out of all the 45 PCA movement components had influence on the yaw rate, $Sim_1$ would be one.

\subsubsection{Dominant DOFs in a given PCA movement component}\label{subsubsec234}

Each PCA movement component that generates a meaningful manoeuvre can be further investigated. Consider feeding the skydiver simulator by a given PCA movement component, i.e. by a given eigenvector $PC_i$ that has 45 elements, each representing a joint rotation DOF. Instead of engaging all DOFs, only an increasing sub-set of these DOFs will be engaged in each simulation, starting from the elements of $PC_i$  with highest absolute values. This way, each element $k$, $k \in [1,45]$ of the body pose is computed at every instant of time as:
\begin{equation}
    pose_i[k] = \begin{cases}
    N_{pose}[k]+\alpha_i(t) \cdot PC_i[k], & k \in k_{eng}\\
    N_{pose}[k], & \text{otherwise}
  \end{cases}
  \label{eq8}
\end{equation}
where $k_{eng}$ is a set of indices of $PC_i$ elements being engaged. 
This will make it possible to determine which DOFs within the studied movement component are most important for performing the manoeuvre associated with this component.  

\subsubsection{Synergies of PCA movement components}\label{subsubsec235}

Synergies of PCA movement components can be determined by activating a combination of two or more components with synthetic periodic control signals that are similar to the signals contained in the relevant rows of matrix in Eq. \eqref{eq2}. PCA control signals can be plotted in the same figure and examined for any apparent correlation between them, such as phase shift. If found, such synergies can be tested in simulation. This allows identifying less dominant components that compensate for undesirable effects more dominant components might have on the overall manoeuvre.  

\subsection{Analysis of movement patterns from a control engineering perspective}\label{subsec24}
\begin{figure*}[h]%
\centering
\includegraphics[width=0.8\textwidth]{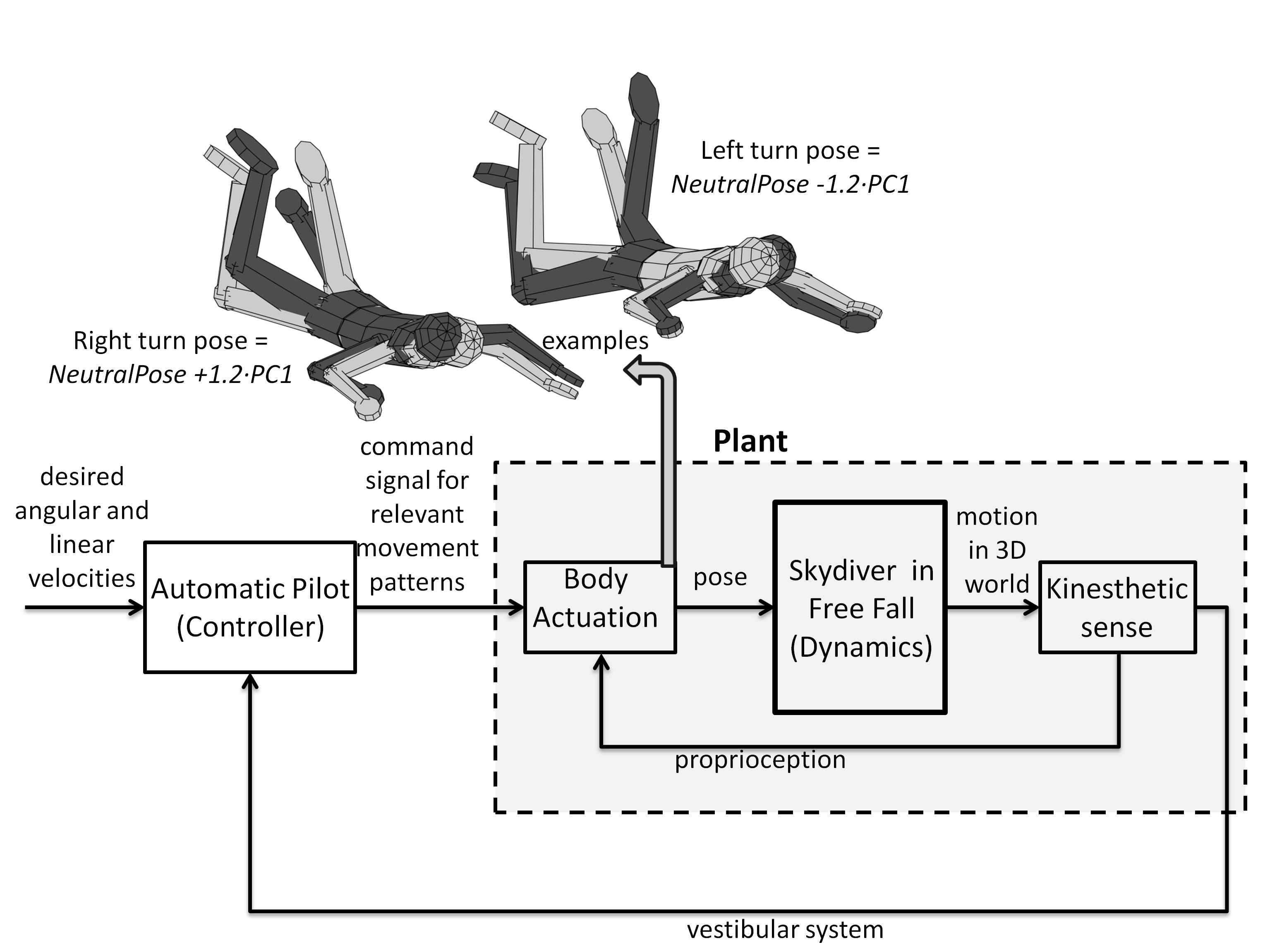}
\caption{Block diagram of a hierarchical closed-loop feedback control system: skydiver performing manoeuvres in free-fall. Body actuation examples show the movement pattern PC1 (the first Principal Pattern extracted from the turning experiment with the Instructor) with control signals  (dark figures) compared with the neutral pose (light figures).}\label{fig2}
\end{figure*}
The key of our method is determining the potential effectiveness of the identified movement components: how well they fulfil their roles, reveal the pitfalls and possible ways to eliminate them. Our method is intended to be a tool for coaches that can analyse why certain posture changes generate faster manoeuvres, performed with less effort, or greater precision; and compute individual modifications for the trainee's movement patterns, which could be naturally produced by the CNS in due course. One of the central problems in the motor learning field is Bernstein's DOFs problem \cite{sport_bernstein1967co}: How does the CNS choose which DOFs to use for a certain movement? We offer in this section a novel insight into this problem applied to manoeuvring in free-fall. In the literature it is usually assumed that movement patterns are solutions of optimizing some cost function. Suggested costs include mechanical energy consumption, joints acceleration/jerk, amount of torques/forces applied by joints/muscles \cite{berret2011evidence}. These cost functions are tested, for example, on arm reaching movements under some disturbances. However, such an approach may not always capture the dynamics of realistic activities. For instance, torques and forces applied by joints and muscles in free-fall are usually very small since all the manoeuvres are executed by the aerodynamic forces and moments, which are generated by only slight changes in the body posture. For these reasons, we offer a new approach to exploration of the DOFs problem and, thus, technique evaluation.

Let’s consider a skydiver performing free-fall manoeuvres from a perspective of automatic control theory. Manoeuvre execution can be represented as a hierarchical closed-loop feedback control system, as shown in Fig. \ref{fig2}. The inner control loop is closed inside the body actuation block, where the command signal, issued for the specific movement patterns, is compared to the feedback from body proprioceptors triggering the necessary adjustments. The outer control loop has a role of an automatic pilot in autonomous systems: an automatic controller that interprets the disparity between the desired and sensed linear and angular velocities in terms of an actuator command. For example, let’s consider the turning manoeuvre discussed in our study case. The desired yaw rate is compared with the actual body yaw rate sensed by the eyes and vestibular system, and according to the disparity the automatic pilot issues a command signal for the pattern $PC_1$. Next, this command is implemented by the body (at a much higher rate than the outer loop) and the resulting joints rotation and posture becomes an input to the equations of motion, that propagate the position, orientation, and motion of the skydiver in a 3D world. In Automatic Control Theory a combination of process and actuator is called plant. As shown in Fig. \ref{fig2}, in our skydiver model, the plant that the automatic pilot has to control includes: actuator - body actuation using a specific movement pattern; process - skydiver free-fall dynamics, and sensing. 

The complexity of an automatic pilot and its performance (how fast and accurate it tracks the desired yaw rate profile) depend on the dynamic characteristics of the plant. When control engineers design an automatic pilot for autonomous platforms they usually have a set of requirements regarding these characteristics and analytical tools for examining the plant and its properties. We believe that the same control considerations are important for a natural automatic pilot built into an athlete’s mind/ body. The body’s natural controller will achieve a better execution of a desired manoeuvre if the plant possesses certain (convenient for control) qualities. The plant, in its turn, strongly depends on the choice of movement patterns that are used for body actuation. Thus, we suggest studying the dynamic characteristics of the plant actuated by a movement pattern under investigation. One of the basic characteristics is the transfer function, which defines the relation between the input and output signal of a system, mathematically defined using Laplace transform \cite{aastrom2010feedback}. For skydiving, the input is a pattern control signal, and the output is a state variable (skydiver’s acceleration/ velocity/ position/ orientation) associated with this pattern. For the turning manoeuvre it will be the transfer function from the pattern angle to yaw rate. In the general case, multiple transfer functions may be constructed: from the input signal of each dominant movement component identified by PCA to each significant state variable. 

An example of such analysis is given in \cite{article}, where we compared two movement patterns for turning obtained from observing two types of students: novice and advanced. Both patterns consisted of only four DOFs (associated with shoulders) but produced plants with very different characteristics. The plant actuated by the ’novice’ pattern had a resonance and anti-resonance pair around the frequencies of a desired bandwidth. It was impossible to design a yaw rate controller for this plant that would satisfy a minimal set of reasonable specifications. This may explain the fact that novice skydivers turn very slowly, whereas initiating a faster turn causes them to lose stability (e.g. flip to back). However, the resonance and anti-resonance pair did not exist in the plant actuated by the pattern of the advanced skydiver who was able to perform fast and stable turns. 
Transfer functions are constructed according to the correlation method \cite{ljung1998system}, while the Skydiving Simulator acts as a virtual spectrum analyser. It is fed with a sequence of poses constructed from the pattern under investigation (the PCA movement component $PC_i$) inputted in a form of a sine wave (Eq. \eqref{eq4})
with different angular frequencies (in our case: $0.01 \leq \omega \leq 100$ rad/s).  
For each frequency, the gain $G(\omega)$ and the phase $\Phi(\omega)$ of the desired transfer function are computed as:
\begin{equation}
\begin{gathered}
G(\omega)=\frac{2\sqrt{y_{c}^2+y_{s}^2}}{A} \\
\Phi(\omega)=arctan\frac{y_{c}}{y_{s}} \\
y_{c}=\int_0^{nT} \Omega(t) \cdot cos(\omega \cdot t) dt \\ 
y_{s}=\int_0^{nT} \Omega(t) \cdot sin(\omega \cdot t) dt \\ 
\end{gathered}
\label{eq9}
\end{equation}
where $A$ is the amplitude of the movement pattern angle, $T=2\pi/\omega$, $n$ is the number of periods taken for analysis. We used $n=10$ cycles, starting after the transient, i.e. after the yaw rate had reached periodic steady state. $\Omega(t)$ 
is the yaw rate of the skydiver computed by the simulator, and the integrals are approximated via the trapezoidal method. Next, a Bode plot of the transfer function is created, where the gain in dB is: $20 \cdot log_{10}G(\omega)$. This procedure is repeated for different amplitudes $A$ of the input signal in order to verify the linear behaviour of the plant in the chosen range of amplitudes. In case of a linear plant all the obtained transfer function values for a given frequency will be identical. 

\section{Results}\label{sec4}

\subsection{Instructor}\label{sec41}
\begin{figure*}[h]%
\centering
\includegraphics[width=1\textwidth]{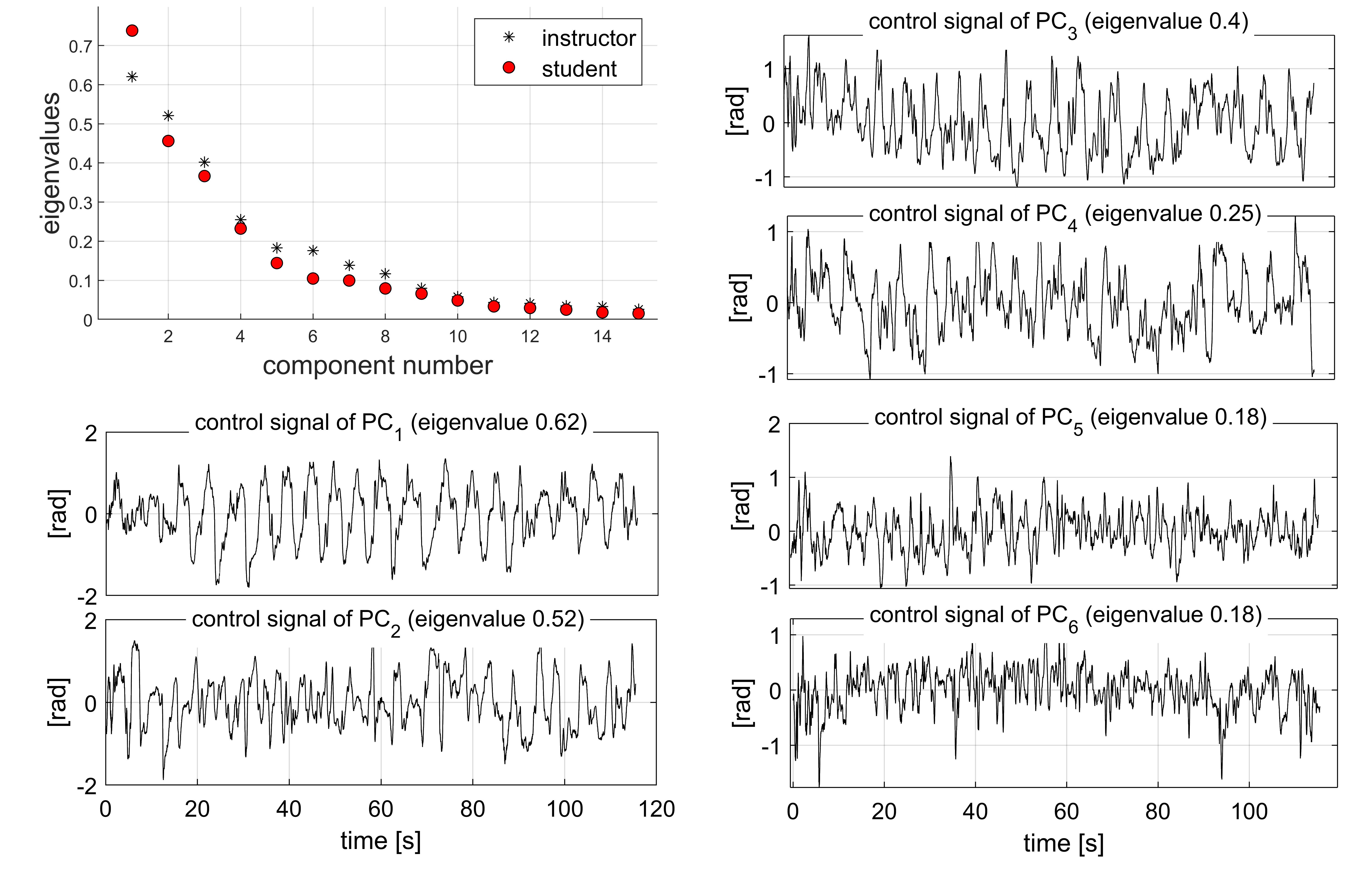}
\caption{Principal Component Analysis of turning: Eigenvalues of the first 15 Principal Components (PCs), and control signals of the first six PCs extracted from experiment with the Instructor. The pose is constructed from PCs according to Eq. \eqref{eq3}}\label{fig3}
\end{figure*}
The PCA of the recorded body postures of the Instructor shows that, similar to other sports (e.g. skiing, see \cite{federolf2014application}), skydiving also requires from our body to produce about 4-8 significant movement components, see Fig. \ref{fig3}. The PCA control signal of the most dominant movement component ($PC_1$) has the same fundamental frequency (0.25 Hz) as the recorded yaw rate. This means that the first PCA movement component is responsible for turning, which was the main objective in this experiment. Other objectives were: keeping distance from all tunnel walls, keeping constant altitude relative to the tunnel floor, and transitioning to each turn in the opposite direction in minimal time. It seems that 4-5 PCA movement components are sufficient for performing the manoeuvre under investigation, since the control signals of the other PCs resemble noise. 
\begin{figure*}[h]%
\centering
\includegraphics[width=0.8\textwidth]{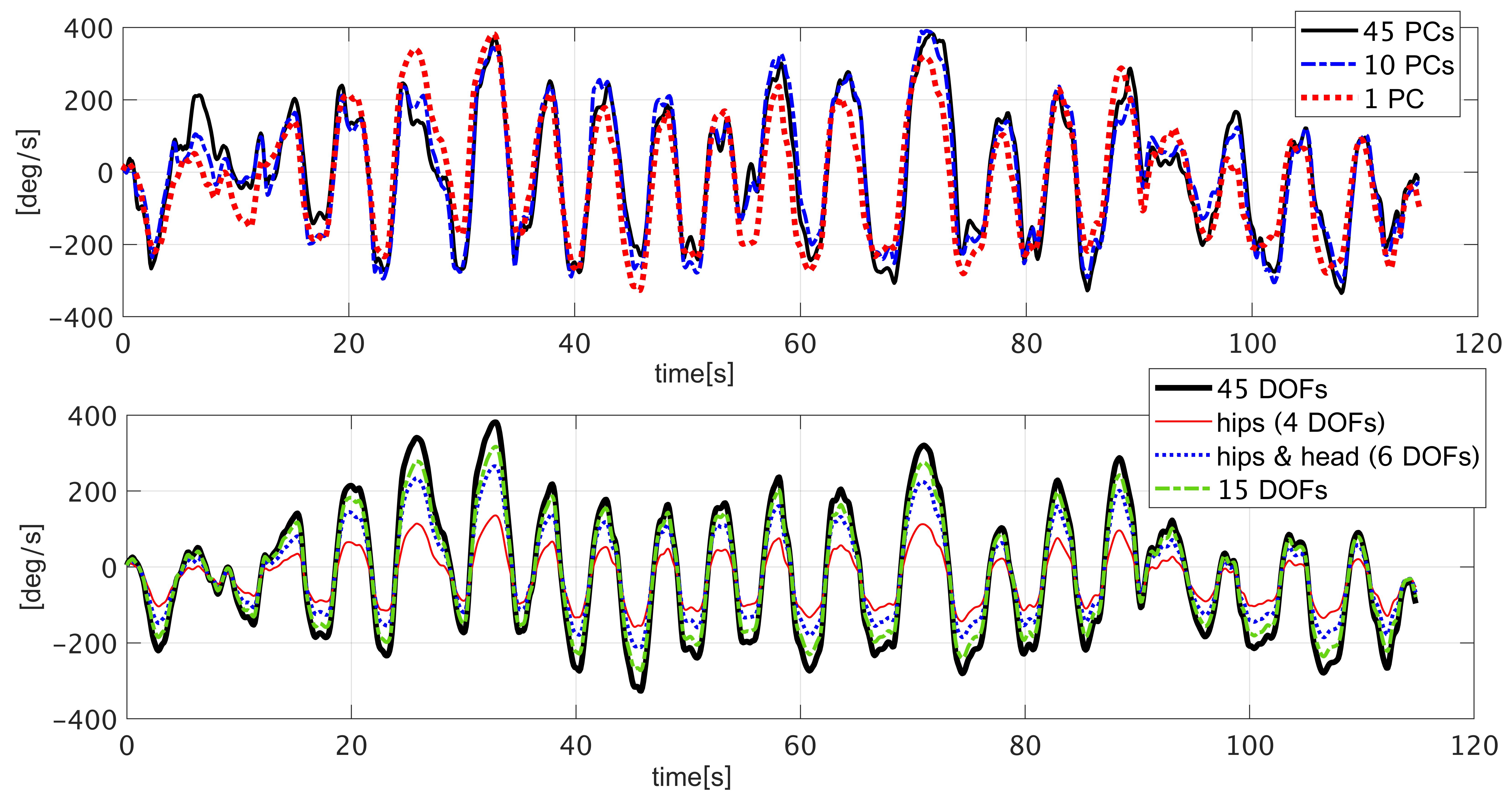}
\caption{Comparison of the yaw rate generated by the Skydiving Simulator using: (top) different amounts of Principal Components (PCs) extracted from experiment with the skydiving instructor; (bottom) $PC_1$ with different amounts of active Degrees-of-Freedom (DOFs), engaged according to Eq. \eqref{eq8} }\label{fig4}
\end{figure*}
From Fig. \ref{fig4} we compute: $Sim_1=0.85$, meaning that one PCA movement component is sufficient to reconstruct 85\% of the yaw rate profile. The highest six values in the eigenvector associated with this component are mostly related to the DOFs of head and hips, see Tab. \ref{tab1} summarizing the values of Euler angles representing the movement component and Fig. \ref{fig1} explaining the coordinate system. 
\begin{figure}[h]%
\centering
\includegraphics[width=0.4\textwidth]{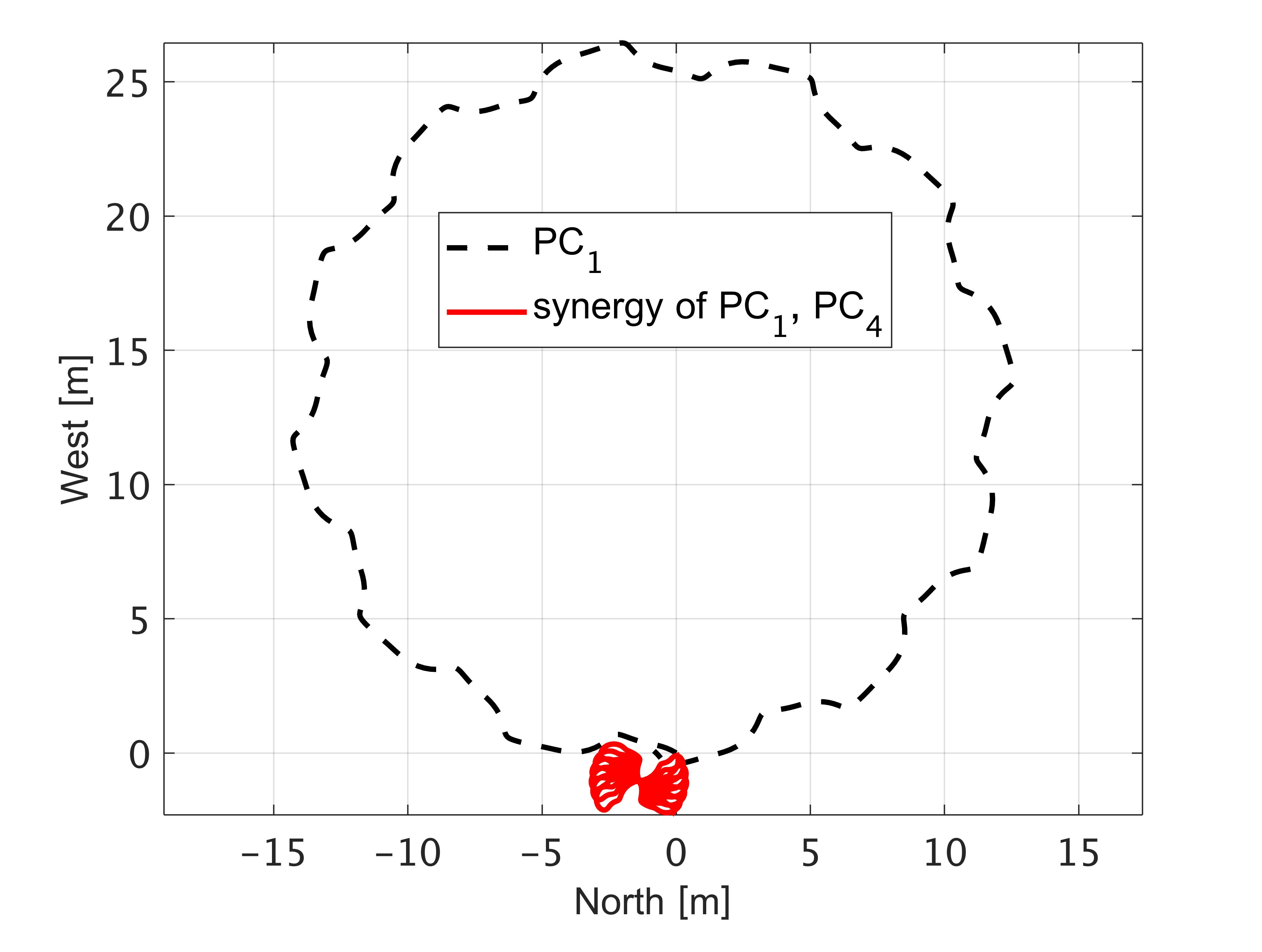}
\caption{Skydiver horizontal position during simulation of a turning manoeuvre using Principal Components (PCs) extracted from experiment with the Instructor }\label{fig5}
\end{figure}
It is known from skydiving experience that hips are very efficient for turning, therefore it is expected that an experienced skydiver engages his hips while turning. The head movement is also expected: we naturally look in the direction of motion. An interesting question, however, is whether the head position is important for reaching the desired yaw rate. As shown in Fig. \ref{fig4}, 84\% of the yaw rate profile ($Sim=0.84$) is reconstructed by engaging only the head and the hips for $PC_1$. However, without engaging the head only 52\% ($Sim=0.52$) of the motion is reconstructed. Thus, the head movement combines two tasks: looking where the body is going, and acting as a significant aerodynamic control surface. Animation of $PC_1$ is shown in Online Resource \ref{res4}. Investigation of the three next dominant PCA movement components shows that:
\begin{itemize}
    \item $PC_2$ provides fall rate adjustments by engaging the arms
    \item $PC_3$ is responsible for stopping the turns by engaging the knees and, to a smaller extent, the shoulders
    \item $PC_4$ prevents orbiting, meaning a horizontal motion in a large circle, induced by $PC_1$
\end{itemize}
Engaging $PC_4$ in addition to $PC_1$ in the following way:
\begin{equation}
    \label{eq10}
     \begin{gathered}
     pose(t)=N_{pose}+
     PC_1\cdot 1.2\cdot sin(\omega\cdot t)+ \\
     PC_4\cdot 0.8\cdot sin(\omega\cdot t+\pi)
     \end{gathered}
\end{equation}
where $\omega=2\cdot \pi \cdot 0.2$ rad/s , compensates for the orbiting effect, see Fig. \ref{fig5}.

\begin{table*}[h]
\begin{center}
\begin{minipage}{\textwidth}
\caption{Degrees-of-Freedom (DOFs) defining movement patterns that generate rotations of the Instructor and the Student}
\renewcommand{\arraystretch}{2}
\begin{tabular*}{\textwidth}[htbp]{l p{0.1cm} c c p{0.2cm} c c c p{0.2cm} c l p{0.2cm} c c c p{0.2cm} c l} 
 \toprule
 & &\multicolumn{2}{l}{Head}& &\multicolumn{3}{l}{Right Shoulder}&  \multicolumn{3}{l}{Right Elbow}& &\multicolumn{3}{l}{Left Shoulder}&  \multicolumn{3}{l}{Left Elbow} \\ 
 Pattern& & $\theta$ & $\psi$&  & $\phi$ & $\theta$ & $\psi$& & $\phi$ & $\psi$&  &$\phi$ & $\theta$ & $\psi$& & $\phi$ & $\psi$  \\ 
 \hline
I-$PC_1$& & 15 &19&  & 9 & -3 & -6& & -25 & 12 &  &  &  & -13&  & 1 & -4\\
\hline
S-$PC_1$& & -21 &-24&  & -5 & 7 & -11& & -5 & 12&  & 7 & 7 & -13& & 6 & 14\\
\hline
S-$PC_6$& & -1& -7& & 9 & -6 & -11 & & -14    & 5 & & 10 &  & 3&  & 20 & 1 \\ 
\end{tabular*}


\begin{tabular*}{\textwidth}[htbp]{ l p{0.05cm} c c p{0.05cm} c c p{0.05cm} c c p{0.05cm} c c p{0.05cm} c c p{0.05cm} c c} 
\toprule
 & &\multicolumn{2}{l}{Right Hip}& &\multicolumn{2}{l}{Right Knee}&  &\multicolumn{2}{l}{Left Hip}& &\multicolumn{2}{l}{Left Knee}&  &\multicolumn{2}{l}{Abdomen}& & \multicolumn{2}{l}{Thorax} \\ 
   
 Pattern& & $\phi$ & $\psi$& & $\phi$ & $\psi$& & $\phi$ &$\psi$&  & $\phi$ & $\psi$ &  &$\theta$ & $\psi$& &$\theta$ & $\psi$ \\ 
 \hline
I-$PC_1$& & -25 & -14& & 7 & -5&  & 20 & -10& &  & -7 &  &2.5 & & & 3 &\\
\hline
S-$PC_1$& & -4 & 6& & 5 & 3& & 4 & 6& &  & 4 &  & -1 & 6 &  &-2 & 7.5 \\
\hline
S-$PC_6$& & -7& -5 & &  & -7& & 17 & -6  & & -20 & & & 1 & 1 &  &1 & 1 \\ 
 \hline
\end{tabular*}
\renewcommand{\arraystretch}{1}
\vspace*{0.5cm}

 \footnotetext{Note: 
The symbols $\phi, \theta, \psi$ in the table head represent Euler angles stated in degrees in order to
facilitate their intuitive interpretation. The values represent the quantity 1 [rad] $\cdot$ $PC$, where $PC$ is the dimensionless movement pattern eigenvector. I-$PC_1$ is the first principal component
extracted from the turning experiment with the Instructor. S-$PC_1$ and S-$PC_6$ are the first and the sixth principal components extracted from the turning experiment with the Student. The DOFs
with absolute values less than 0.5 [deg] are not shown (angles of wrists and ankles, $\phi$ of thorax
abdomen and head, and $\theta$ of elbows, knees and hips).
 }
\label{tab1}
\end{minipage}
\end{center}
\end{table*}

\subsection{Student}\label{sec42}
\begin{figure}[h]%
\centering
\includegraphics[width=0.5\textwidth]{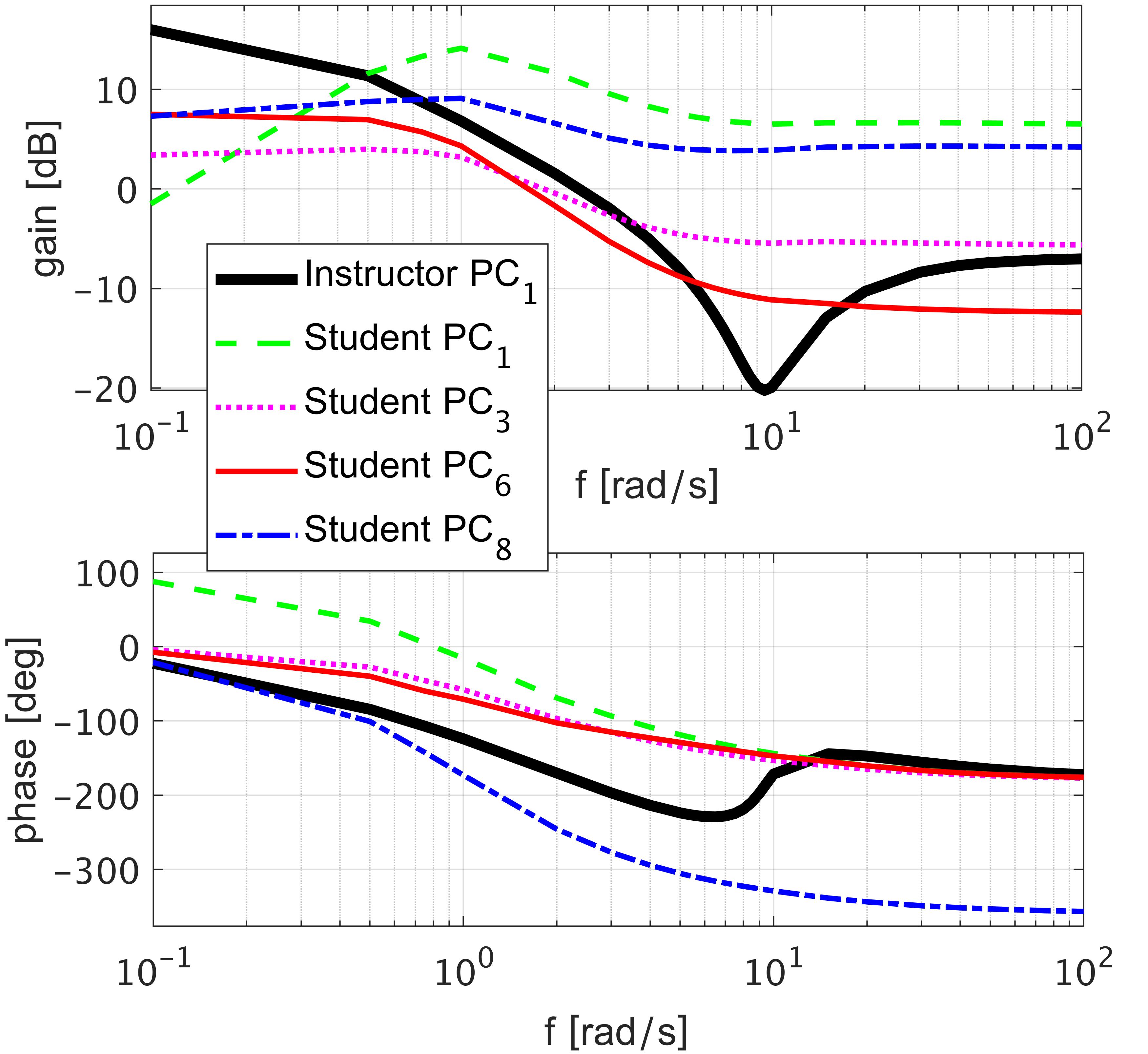}
\caption{Comparison of the pattern angle to yaw rate transfer functions of the Instructor and the Student. The movement patterns used for the transfer function generation are the Principal Components (PCs) extracted from experiments }\label{fig6}
\end{figure}
The Student exhibited many different PCA movement components incorporated for turning (10-15 as opposed to one turning component exploited by the Instructor; the control signals for the six most dominant components are shown in Online Resource \ref{res5}). Most of these components possess serious pitfalls: are coupled with a horizontal/vertical motion component, provide a turn only in one direction, and require very large amplitude of a control signal. The yaw rate profile of the experiment is closely reconstructed by the first nine PCA movement components ($Sim_9=0.87$, see Online Resource \ref{res6}).  Thus, the Student's movement repertoire includes a greater amount of less efficient coordination patterns. This is verified by the Bode diagram in Fig. \ref{fig6} comparing the Instructor's turning pattern with the student ones, animated in Online Resources \ref{res7}-\ref{res10}.  The Student's transfer functions from turning patterns to yaw rate lack gain at low frequencies (making it hard to generate a turn and keep high turning rate), and lack attenuation at high frequencies (making it hard to stabilize the closed loop and deal with disturbances and noise). These are the reasons the Student reported that he had to constantly fight the turbulence of the tunnel airflow and found it hard to generate the movement. The Instructor's transfer function has a high gain at low frequencies providing the efficiency of rotations, and a gain reduction of about 20 dB per a decade starting from around 1 rad/s, providing a disturbance attenuation at high frequencies and replicating dynamics of an integrator in the innermost (proprioception) loop, recall the control block diagram in Fig. \ref{fig2}. 
The higher level loop, which tracks the yaw rate, includes the dynamics of the closed proprioception loop: replicating a low pass filter, what is also seen in the phase plot in Fig. \ref{fig6}.

In this way, the transfer function of the plant possesses the desired characteristics of the open yaw rate tracking loop, which can thus consist of the plant and a proportional controller. In other words, the control signal driving the Instructor's turning pattern can be simply the scaled yaw rate error. In contrast, any of the Student's plants will be either inefficient or unstable under proportional control. It is possible to design a controller with high order dynamics in order to compensate the plant pitfalls and achieve a desired closed loop behaviour, but it is unlikely that such a controller can be implemented by a human motor system. 

From numerous experiments with various aerial manoeuvres we may hypothesise that humans can not implement complex dynamic controllers. Instead, the human body develops such movement patterns that produce an ideal plant for practised manoeuvres, enabling athletes to operate in a closed loop via a simple control law. 

\subsection{Proposing improvement of Student's technique with simulation}\label{sec43}
\begin{figure}[h]%
\centering
\includegraphics[width=0.48\textwidth]{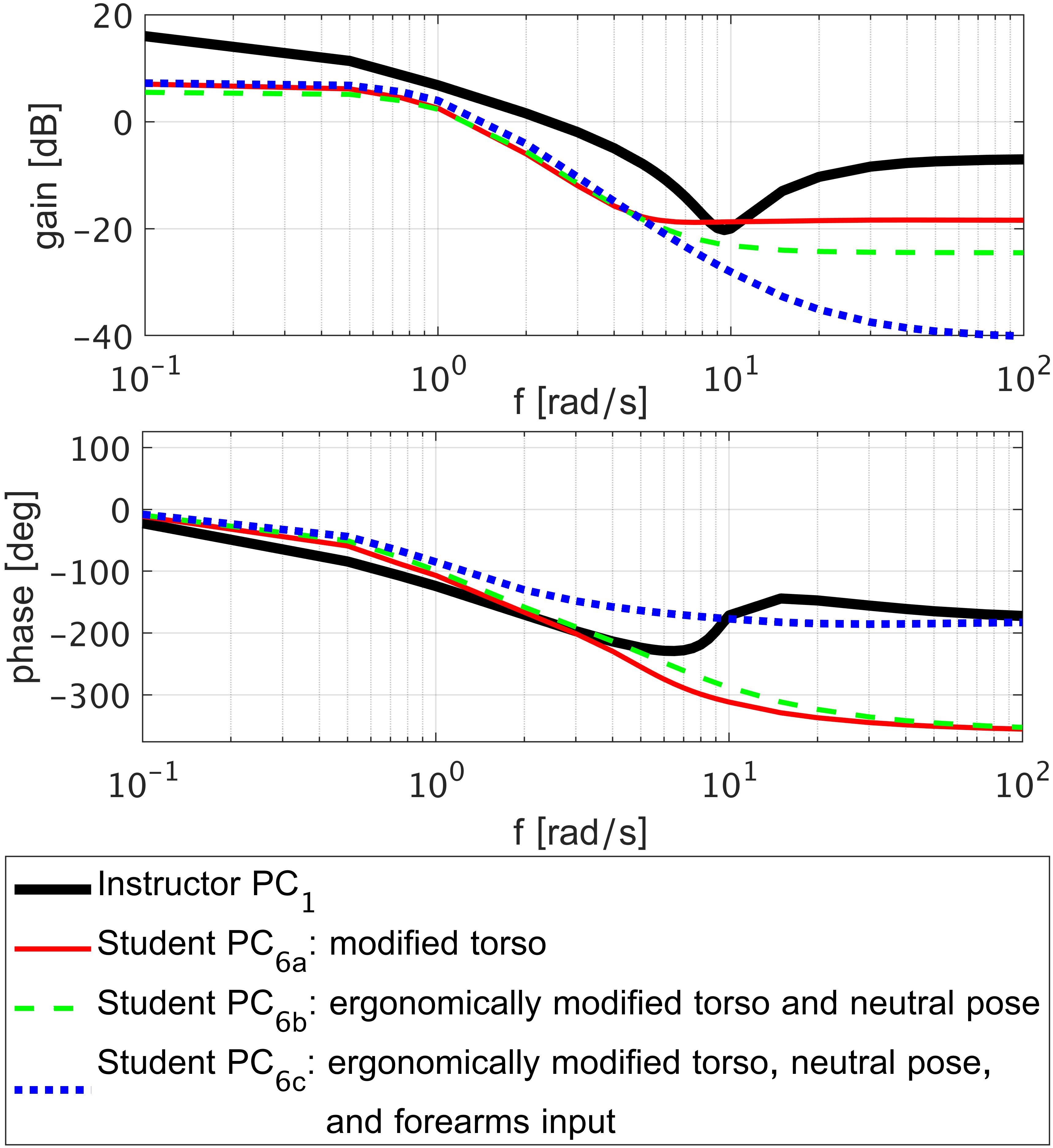}
\caption{Comparison of the pattern angle to yaw rate transfer functions of the Instructor and the Student. The movement patterns used for the transfer function generation are the first Principal Components (PC) of the Instructor and the sixth PC of the Student, modified in various ways }\label{fig7}
\end{figure}
The above analysis of the Student's pattern angles to yaw rate transfer functions showed that a significant improvement of the student's technique will follow from increasing the dynamic stiffness property of his plant. Dynamic stiffness, or impedance, is the ability of the actuator to resist an external oscillatory load \cite{wang2015commercial,winters1988analysis}. Hence, we seek to increase the student's ability to reject disturbances at high frequencies, such as the turbulence of the wind tunnel air. In this section we introduce a possible modification of one of the student's turning patterns, so that this goal is achieved. 

The chosen pattern is the PCA movement component $PC_6$, since its corresponding transfer function has less pitfalls relative to other options. $PC_6$ is not very dominant (its eigenvalue is 0.11), which means it was used for only a few of the performed turns. The reason, probably, is that this coordination pattern has been formed only recently, it still cohabits with other turning patterns in the movement repertoire, and is often overpowered by ’old habits’. 

Such a dominant property as dynamic stiffness is most likely related to the major aerodynamic surface of the body – the torso. From Tab. \ref{tab1} it can be seen that the torso is activated differently by the Instructor and the Student during turning. The Instructor exhibits lateral tilt, whereas the Student exhibits axial rotation, especially this is pronounced in his most dominant PCA movement component $PC_1$. Notice (from Tab. \ref{tab1}) that axial rotation in $PC_6$  becomes much smaller, and even some tilt appears. Modifying $PC_6$ so that the torso tilts instead of rotating (abdomen $\theta=2.5$ degrees, $\psi=0$; thorax $\theta=0.5$ degrees, $\psi=0$) significantly changes the pattern angle to yaw rate transfer function, bringing it very close to the Instructor's one, see the Bode plot in Fig. \ref{fig7}  labelled '$PC_{6a}$ modified torso'. Just a couple of degrees of torso tilt make a great difference. We called this effect \textit{the key DOF} - DOF that has the most impact on shaping the dynamic response. 

\begin{figure}[h]%
\centering
\includegraphics[width=0.5\textwidth]{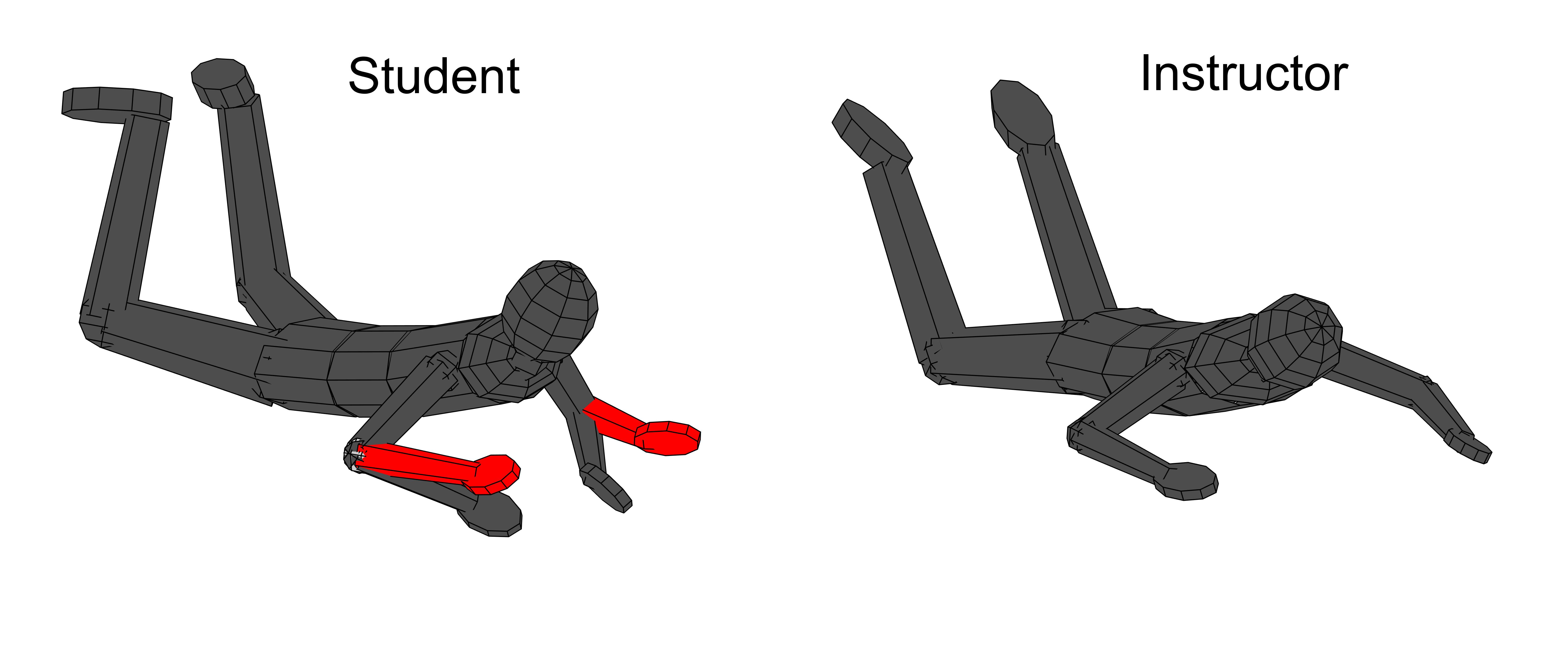}
\caption{Comparison of the neutral pose of the Instructor and the Student. Modification of the Student's forearms position (red) }\label{fig8}
\end{figure}
\begin{figure*}[h]%
\centering
\includegraphics[width=0.9\textwidth]{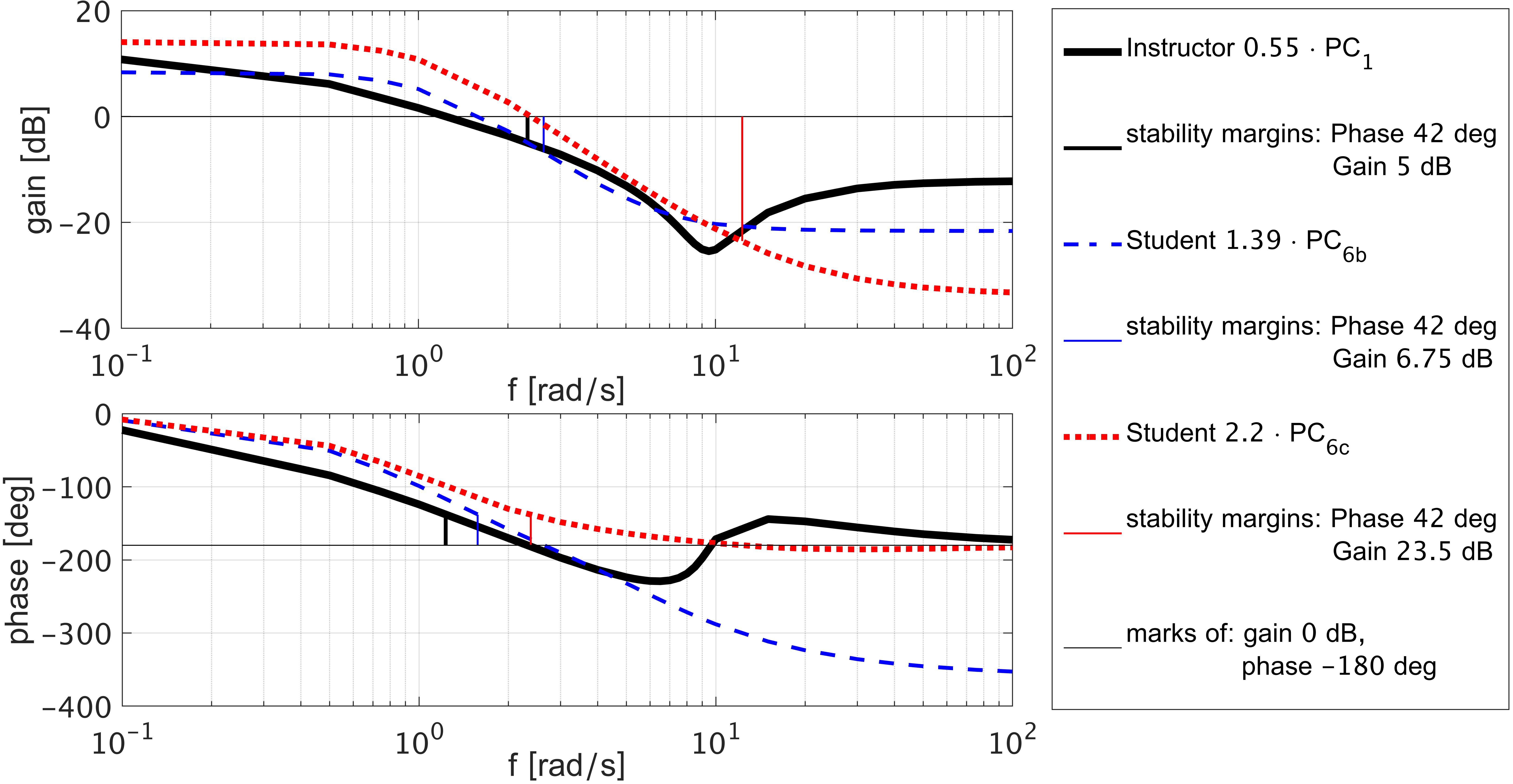}
\caption{Comparison of the open yaw rate loop transfer functions of the Instructor and the Student. The movement patterns used for the transfer function generation are the first Principal Components (PC) of the Instructor and the sixth PC of the Student, modified in torso ($PC_{6b}$), and modified in torso and elbow joint ($PC_{6c}$). Both Student plants include a neutral pose modification }\label{fig9}
\end{figure*}
However, in the introduced modification the thorax joint tilt is smaller than the abdomen joint tilt, what is opposed to the torso actuation observed in the instructor. This difference might be caused by differences in body parameters, however, prior to experimental verification the reason remains uncertain and might be ergonomic. Therefore, in case a more comfortable torso movement implies tilting thorax more than the abdomen, the Student is offered a second option. One change in the Student's neutral pose provides an option to actuate his torso similar to the Instructor (abdomen $\theta=1.5$ degrees, thorax $\theta=2$ degrees), while the transfer function under investigation remains close to the one of the Instructor (see Fig. \ref{fig7}, Bode plot labelled '$PC_{6b}$ ergonomically modified torso and neutral pose'). This change is associated with the position of the forearms: in the Student's recorded neutral pose the forearms are too far down (relative to the Instructor's neutral pose and also to the standardly accepted neutral pose).  This is fixed by decreasing the shoulder's internal rotation (by 35 degrees), as shown in Fig. \ref{fig8}. Notice that $PC_6$  must be normalized after modification for computation of the transfer function according to Eqs. \ref{eq4} and \ref{eq9}. 

It is possible to further improve the transfer function under investigation by increasing the phase around the current bandwidth frequency. This will allow using a proportional controller with a larger gain and extending the bandwidth, thus performing faster turns with faster transitions between different turning directions. Such an improvement can be achieved by modifying, in addition to the DOFs mentioned above, the flexion of the elbows: both elbows flexion values $\phi$ are multiplied by 1.7. The resulting transfer function is labelled in Fig. \ref{fig7} as '$PC_{6c}$, ergonomically modified torso, neutral pose, and forearms input'. 

The Student's plant modified as proposed above can be easily controlled using a proportional controller, as shown in Fig. \ref{fig9}. Moreover, the latter modification allows achieving a better gain margin (23.5 dB) than that of the Instructor's open control loop (5 dB). 

In this way, rather than mimicking the Instructor's movement pattern our technique improvement method is aimed at shaping the Student's plant. This process, however, requires to decide, first of all, on design specifications: desired dynamic properties of the plant under investigation. In the next sub-section we address manoeuvres that involve contradicting dynamic characteristics. The rotations performed at a competitive level require to shift the stability-agility trade-off towards the latter, in order to produce the fastest and most accurate rotations a human body is capable of.

\subsection{Elite Skydiver}\label{sec43}
\begin{figure}[h]
\centering
\includegraphics[width=0.5\textwidth]{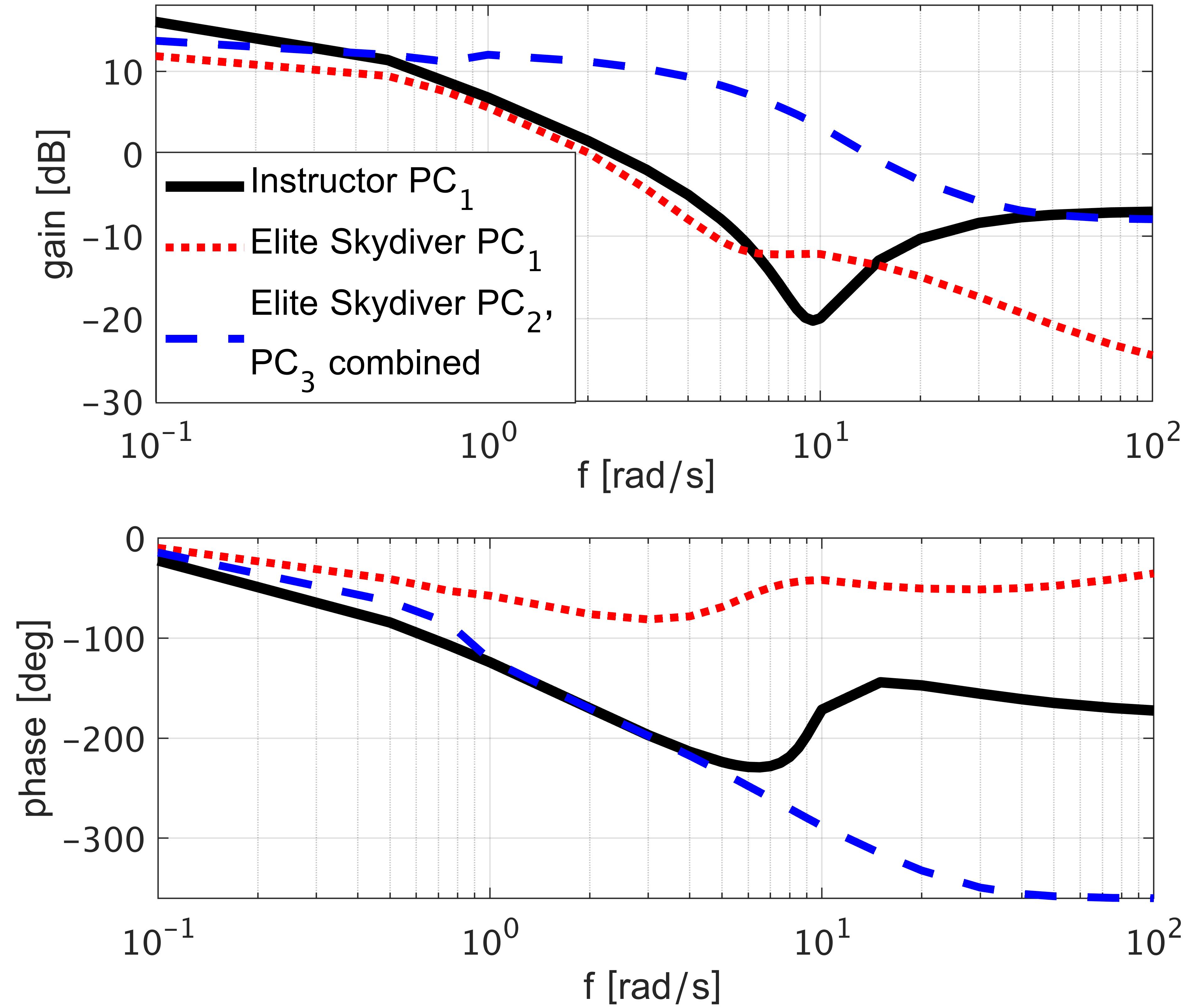}
\caption{Comparison of the pattern angle to yaw rate transfer functions of the Instructor and the Elite Skydiver. The movement patterns used for the transfer function generation are the Principal Components (PCs) extracted from experiments. A combination of $PC_2$ and $PC_3$ is defined in Eq. \eqref{eq11}.
}
\label{fig10}
\end{figure}

\begin{table*}[h]
\begin{center}
\begin{minipage}{\textwidth}
\begin{tabular*}{\textwidth}[htbp]{l p{0.1cm} c c c p{0.2cm} c c c p{0.2cm} c c c p{0.2cm} c c c} 
 \toprule
 &  & \multicolumn{3}{l}{Right Shoulder}&  & \multicolumn{3}{l}{Right Elbow}&  &\multicolumn{3}{l}{Left Shoulder}&  &\multicolumn{3}{l}{Left Elbow} \\ 
 Pattern& &  $\phi$ & $\theta$ & $\psi$& & $\phi$ & $\theta$& $\psi$& &  $\phi$ & $\theta$ & $\psi$& & $\phi$ & $\theta$ & $\psi$ \\ 
 \hline
$PC_1$ & &  4 & -10 & -9&  & -10 & 10 & 4 &  & -7  & -7 & -8&  & 20 & 11 & 11\\
\hline
$PC_2$ & &  -6 & -2 & 4&  & 13 &  &-9  &  &   & 2 &  &  & 20 &  &18  \\ 
\hline
$PC_3$ & &   & -2 &  &  & 20 &  & -18 &  & -6  & 2 & -4&  & 13 &  &9  \\ 
\end{tabular*}


\begin{tabular*}{\textwidth}[htbp]{ l p{0.1cm} c c c p{0.2cm} c c c p{0.2cm} c c c p{0.2cm} c c c } 
\toprule
 & &\multicolumn{3}{l}{Right Hip}& &\multicolumn{3}{l}{Right Knee}&  &\multicolumn{3}{l}{Left Hip}& &\multicolumn{3}{l}{Left Knee}  \\ 
   
 Pattern& & $\phi$ & $\theta$ & $\psi$& & $\phi$ &$\theta$& $\psi$& & $\phi$ & $\theta$ & $\psi$&  & $\phi$ &$\theta$& $\psi$  \\ 
 \hline

$PC_1$& & -6 &-3 & & &  & & -2& & 7 & -6& -2& & -7 & & -2  \\
\hline
$PC_2$& & 12 &17 & & & -18& 4 & & & -5 & & 3& & 2 & 2 & 7  \\ 
 \hline
$PC_3$& & -5 & & -3& & 2 &-2& -7& & 12 &-17 & & & -18 & -4&  \\ 
\end{tabular*}

\begin{tabular*}{\textwidth}[htbp]{ l p{0.1cm} c c p{0.2cm} c c p{0.2cm} c c p{0.2cm} c c p{0.2cm} c c } 
\toprule
 & &\multicolumn{2}{l}{Head}& &\multicolumn{2}{l}{Abdomen}&  &\multicolumn{2}{l}{Thorax}& &\multicolumn{2}{l}{Right Ankle} & 
 &\multicolumn{2}{l}{Left Ankle} \\ 
   
 Pattern& & $\theta$ & $\psi$ & &$\theta$ & $\psi$& & $\theta$ & $\psi$ & &$\theta$ &$\psi$& & $\theta$ & $\psi$ \\ 
\hline
$PC_1$& & -17 & -13& &  &-0.5 & &  & -0.5& & -7 & 11 &  & -8 &9 \\
\hline
$PC_2$& & -10 & -7& & -1.4 &-0.6 & & -2 &-0.5 & & 5 &-19  &  &18  &-11 \\ 
 \hline
$PC_3$& & 10 & 7& & 1.4 & 0.6& & 2 &0.5 & &-18  &11  &  & -5 &19 \\ 
 \hline
\end{tabular*}
\caption{Degrees-of-Freedom (DOFs) defining movement patterns that generate rotations of the Elite Skydiver.}


\footnotetext{Note: The symbols $\phi, \theta, \psi$ in the table head represent Euler angles stated in degrees in order to
facilitate their intuitive interpretation. The values represent the quantity $1 [rad] \cdot PC$, where $PC$ is the dimensionless movement pattern eigenvector. $PC_1$, $PC_2$, $PC_3$ are the first three Principal Components
extracted from the turning experiment with the Elite Skydiver. The DOFs
with absolute values less than 0.5 [deg] are not shown.
 }
\label{tab2}
\end{minipage}
\end{center}
\end{table*}
By the means of PCA three movement components producing  turning were extracted from  the experiment with the Elite Skydiver. See Online Resource \ref{res11}- \ref{res13} for the animation of these patterns, and Online Resource \ref{res14} for the PCA details and reconstruction of the measured yaw rate profile in simulation. The first Principle Component produces a plant that can be stabilized with a  proportional controller, as in the case of the Instructor rotations. Moreover, it  has better phase characteristics, allowing  for very large phase and gain margins, since this system is Minimum Phase, see Fig. \ref{fig10}. It is hence possible to control this plant with a high gain proportional controller. However, we must bear in mind that all human joints have limitations, therefore, high gain will improve performance only until the actuation limits are reached. Nevertheless, it is advantageous to be able to utilize the whole range of movement granted by the body flexibility. 

The second and third Principal Components produce turns only in one direction: right and left, respectively. This is due to body engineering: these patterns, as opposed to $PC_1$, involve an asymmetrical engagement of the hips DOFs, see Tab. \ref{tab2}. In $PC_1$ these DOFs are used symmetrically with a small range of motion: the right hip's flexion and abduction is used along with the same magnitude of left hip's extension and adduction in order to turn left, and vice-a-versa. In $PC_2$ and $PC_3$ only one hip is used for turning in each direction: $PC_2$ generates a right turn using right hip extension and abduction, while $PC_3$ generates a left turn using left hip extension and abduction. Thus, for control purposes these two movement components can be combined, such that either $PC_2$ or $PC_3$ is activated depending on the sign of the control signal $\alpha(t)$, as shown in Eq. \eqref{eq11}.

By simulating the skydiver motion caused by a sine input driving Eq. \eqref{eq11} we can see that movement components $PC_2$, $PC_3$ invoke a different mechanism of turning than that initiated by $PC_1$ of both skydivers. When $PC_1$ is applied (see Online Resource \ref{res15} for simulation recording), the aerodynamic forces acting on arms and legs acquire components that generate a yaw moment depending on lever arms: distances of those limbs from the centre of gravity. 
\begin{equation}
\begin{gathered}
    pose(t)= N_{pose}+ 
\begin{cases}
    \lvert \alpha(t) \rvert \cdot PC_2, & \text{if } \alpha(t)<0\\
    \alpha(t) \cdot PC_3,              & \text{otherwise}
\end{cases}
\end{gathered}
\label{eq11}
\end{equation}

\begin{figure}[h]
    \centering
     \includegraphics[width=0.35\textwidth]{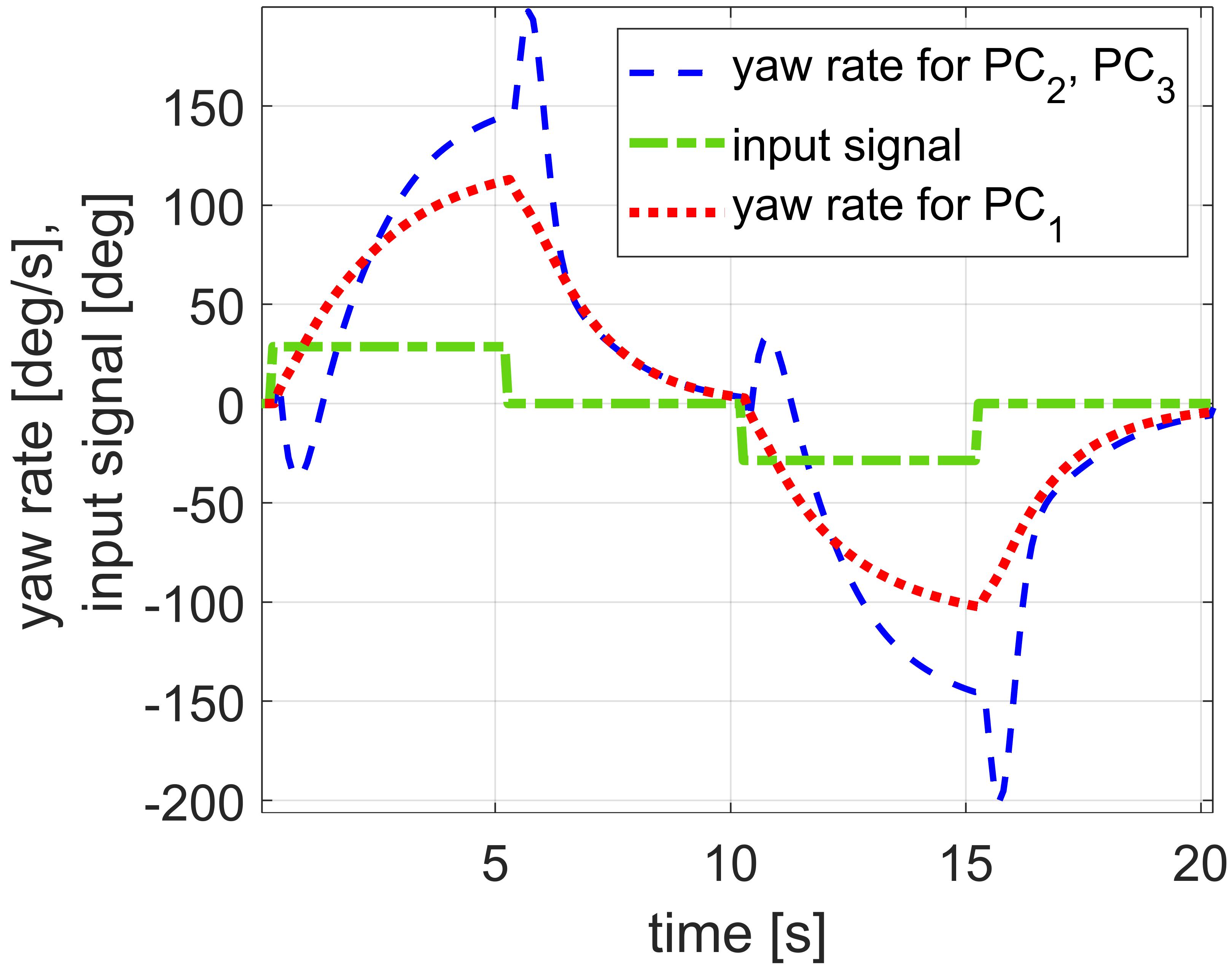}
\caption{Open loop response to the step input signal driving the movement components $PC_1$, $PC_2$, and $PC_3$ extracted from the rotations experiment with the Elite Skydiver. The pose in the case of $PC_2$, $PC_3$ combination is defined according to Eq. \eqref{eq11}}
\label{fig11}
\end{figure}

\begin{figure}[h]
    \centering
     \includegraphics[width=0.35\textwidth]{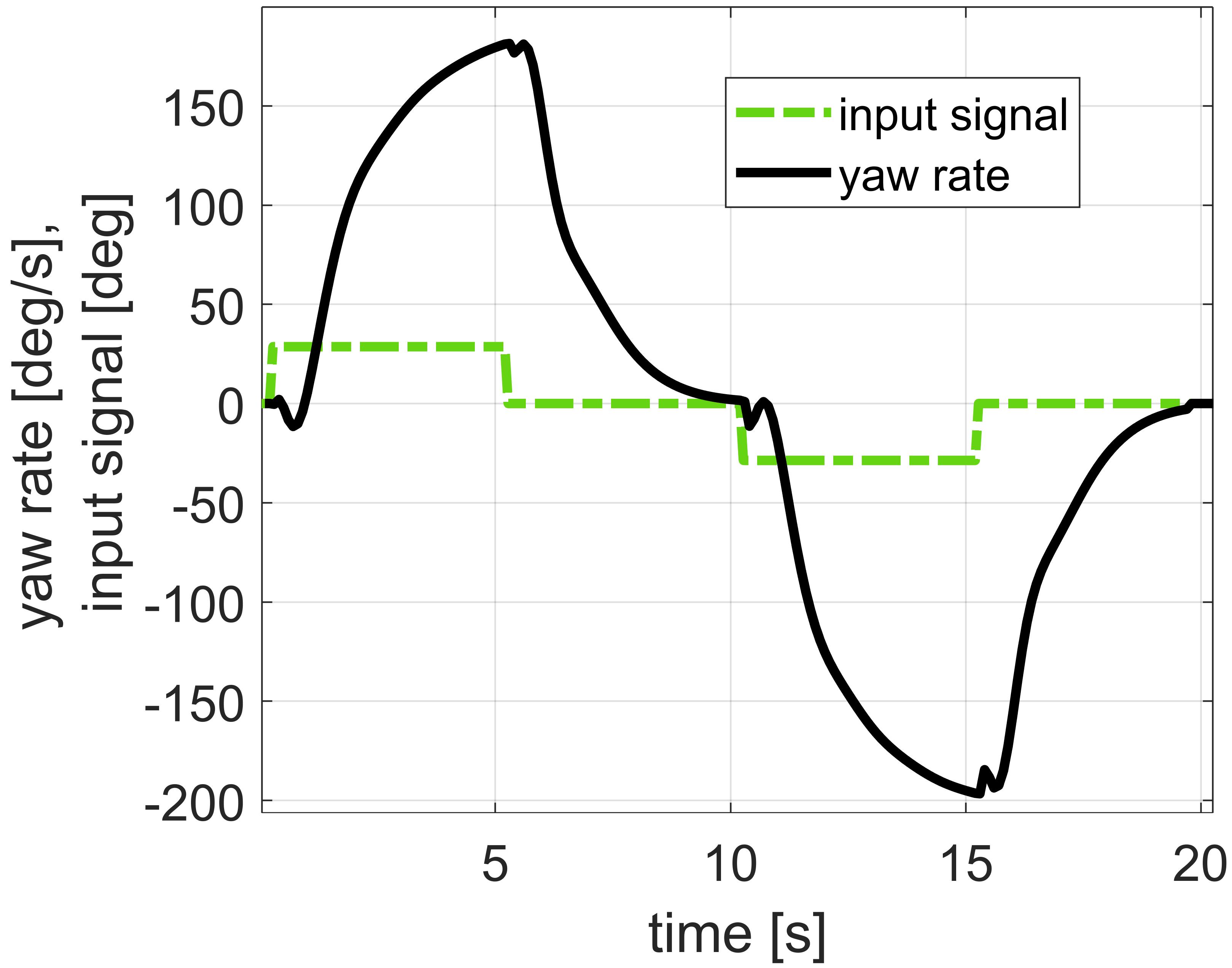}
\caption{Open loop response to the step input signal driving the movement component $PC_1$ extracted from the rotations experiment with the Instructor}
\label{fig12}
\end{figure}

In contrast (see Online Resource \ref{res16}), application of $PC_2$, $PC_3$ causes the roll movement of the upper body, thus creating an angle between the torso and the airflow. This generates a large aerodynamic force responsible for the yaw moment, since torso has the largest surface area of all body segments. During the initial roll motion the body slightly rotates in the opposite direction, which indicates that this plant is Non Minimum Phase (NMP), see Fig. \ref{fig11}. Notice from the phase diagram on Fig. \ref{fig10} that the Instructor's plant is also NMP, however this behaviour is less pronounced, as also seen from the step response on Fig. \ref{fig12}.

Additionally, the transfer function, associated with this plant, has high gains for frequencies up to 10 rad/s, whereas the phase at low frequencies is similar to that of the plant of the Instructor  (see Fig. \ref{fig10}). 
This means that if the movement components $PC_2$, $PC_3$ are engaged with large control inputs, assuming a proportional controller whose gain is larger than about 0.3, the closed loop will be unstable. As it was mentioned in Sect. \ref{sec42},  humans are unlikely to be able to implement a dynamic controller. In the experiment, however, amplitudes reaching 1 rad were observed. 

It seems, therefore, that the Elite Skydiver engages the movement components $PC_2$, $PC_3$ in  open loop: when the right turn is desired $PC_2$ is engaged proportional to the desired angular acceleration, in order to 'throw' the body into a turn, and as it turns, $PC_1$ is used in a closed loop to adjust the turning rate to a desired profile. In order to examine this observation, such a control strategy was implemented in simulation. A sine yaw rate profile was tracked, and the tracking performance achieved by a controller based only on $PC_1$ was compared to the results obtained utilizing a combination of the three movement components.

\begin{figure}[h]
\centering
\begin{subfigure}{.5\textwidth}
  \centering
\includegraphics[width=0.8\textwidth]{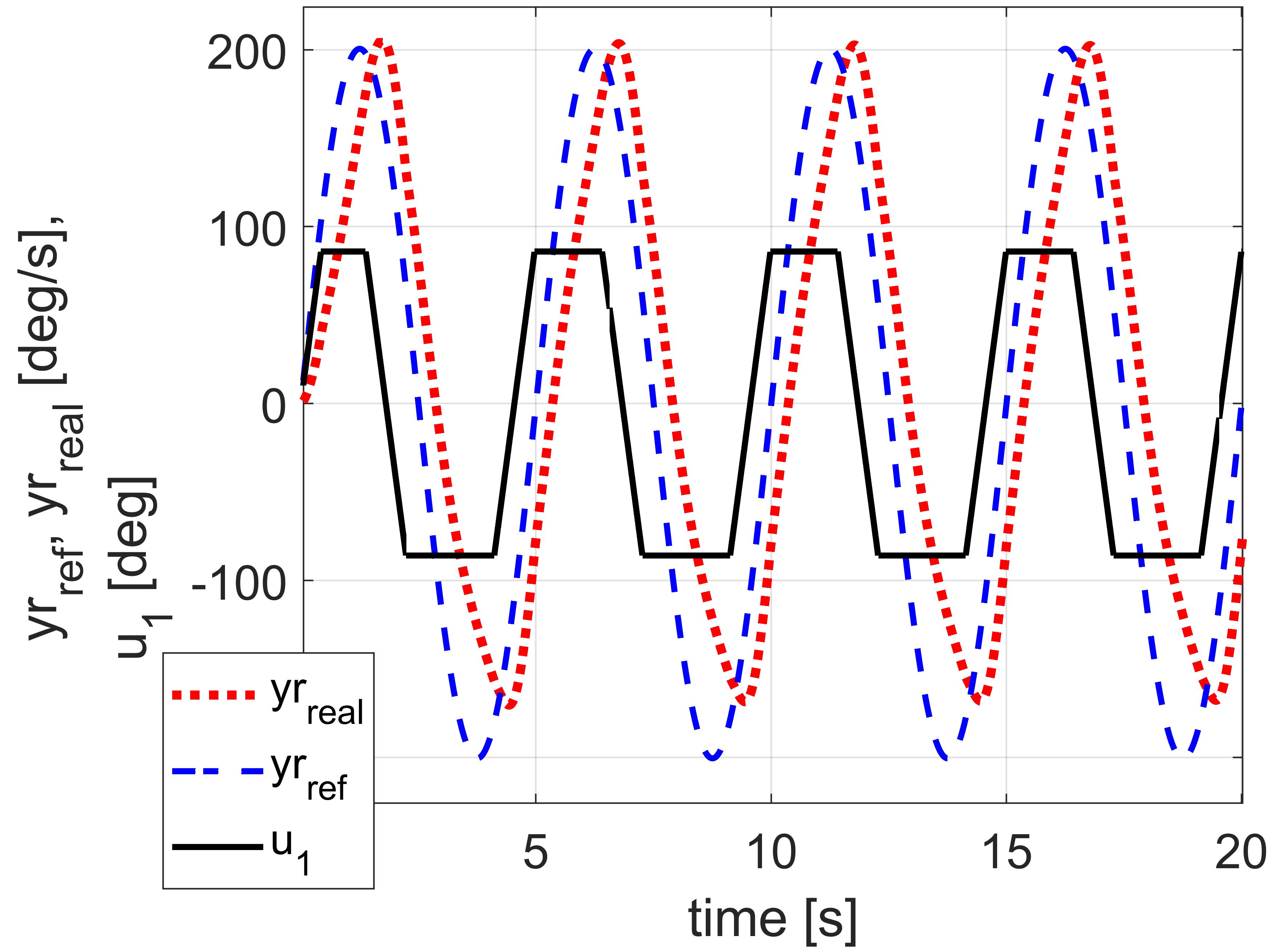}
\caption{pose given in Eq. \eqref{eq14} }
\label{fig13}
\end{subfigure}%

\begin{subfigure}{.5\textwidth}
  \centering
\includegraphics[width=0.8\textwidth]{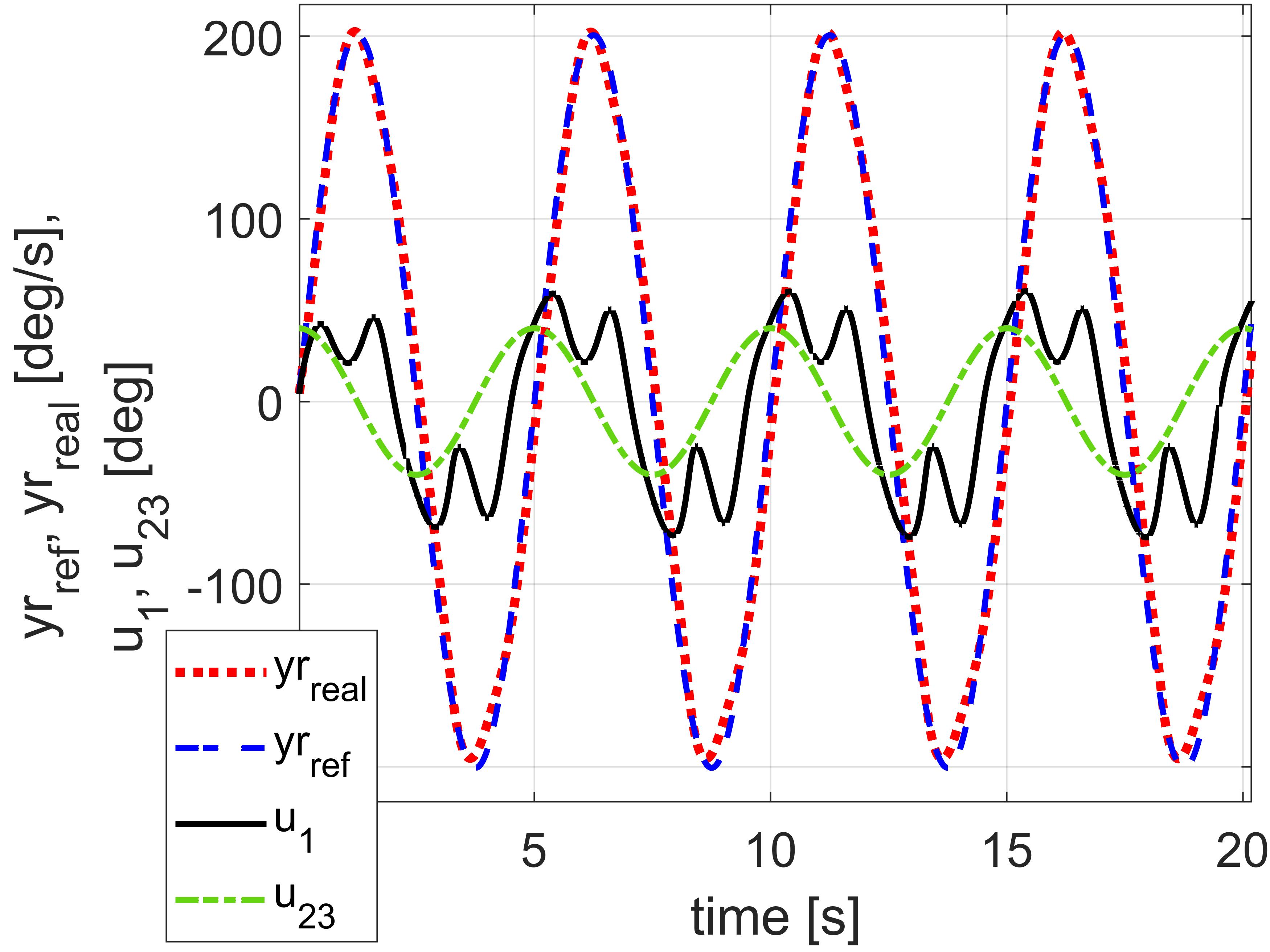}
\caption{pose given in Eq. \eqref{eq17} }
\label{fig14}
\end{subfigure}
\caption{Simulation of tracking a sine yaw rate profile in a closed loop. The movement components $PC_1$, $PC_2$, $PC_3$, extracted from the rotations experiment with the Elite Skydiver, are used for body actuation }
\end{figure}

The yaw rate reference signal was:
\begin{equation}
\Omega_{ref}(t)=3.5\cdot sin(2\cdot \pi \cdot 0.2 \cdot t)
\label{eq12}
\end{equation}
where $t$ is the simulation time. The controller utilizing $PC_1$ for body actuation had  proportional and feedforward parts:
\begin{equation}
\alpha_1(t)=2.5 \cdot (\Omega_{ref}(t)-\Omega(t)) + 0.15 \cdot \Omega_{ref}(t)
\label{eq13}
\end{equation}
where $\Omega(t)$ [rad/sec] is the yaw rate of a virtual skydiver in simulation, and $\alpha_1(t)$ [rad] is the control signal, i.e. the angle of the movement component $PC_1$, such that the skydiver's pose at each instant of time is defined as:
\begin{equation}
pose(t)=N_{pose} + \alpha_1(t) \cdot PC_1
\label{eq14}
\end{equation}
It can be seen from Fig. \ref{fig13} that the desired yaw rate profile can not be tracked accurately due to actuation limitations: 
\begin{equation}
\begin{gathered}
max(\lvert \alpha_1(t) \rvert)=1.5 \quad [rad] \\
max\left( \left \lvert \frac{d\alpha_1(t)}{dt} \right \rvert \right)=3.5 \quad [rad/s]
\end{gathered}
\label{eq15}
\end{equation}
In other words, the movement component $PC_1$ does not allow to change the turning rate so fast. The delay in tracking the desired yaw rate profile is 0.5 s. Utilizing the movement components $PC_2$ and $PC_3$  significantly improves the performance, as shown in Fig. \ref{fig14}.  The feedforward control signal driving the combination of $PC_2$, $PC_3$ is 
\begin{equation}
\alpha_{23}(t)=0.16 \cdot \frac{d(\Omega_{ref}(t))}{dt}
\label{eq16}
\end{equation}
The pose at each instant of time is defined as:
\begin{equation}
\begin{gathered}
    pose(t)= N_{pose}+\\
\begin{cases}
    \alpha_1(t)\cdot PC_1+\lvert \alpha_{23}(t) \rvert \cdot PC_2, & \text{if }  \alpha_{23}(t)<0\\
   \alpha_1(t)\cdot PC_1+\alpha_{23}(t)\cdot PC_3,              & \text{otherwise}
\end{cases}
\end{gathered}
\label{eq17}
\end{equation}
where $\alpha_1(t)$ is according to Eq. \eqref{eq13}. This control strategy, which we termed the \textit{Superposition}, allows to reduce the tracking delay to 0.05 s and reconstruct in simulation the fast turns observed in the experiment, see Fig. \ref{fig14}.

We thus hypothesize that for each specific manoeuvre performed at a competition level athletes develop a 'high performance' pattern (such as $PC_2,\, PC_3$), what may be analogous to the flight dynamics of high performance aircraft. Athletes have to learn from practice how to engage this pattern during the manoeuvre execution, along with the closed loop control of the manoeuvre. 


\begin{figure}[h]
\centering
\includegraphics[width=0.5\textwidth]{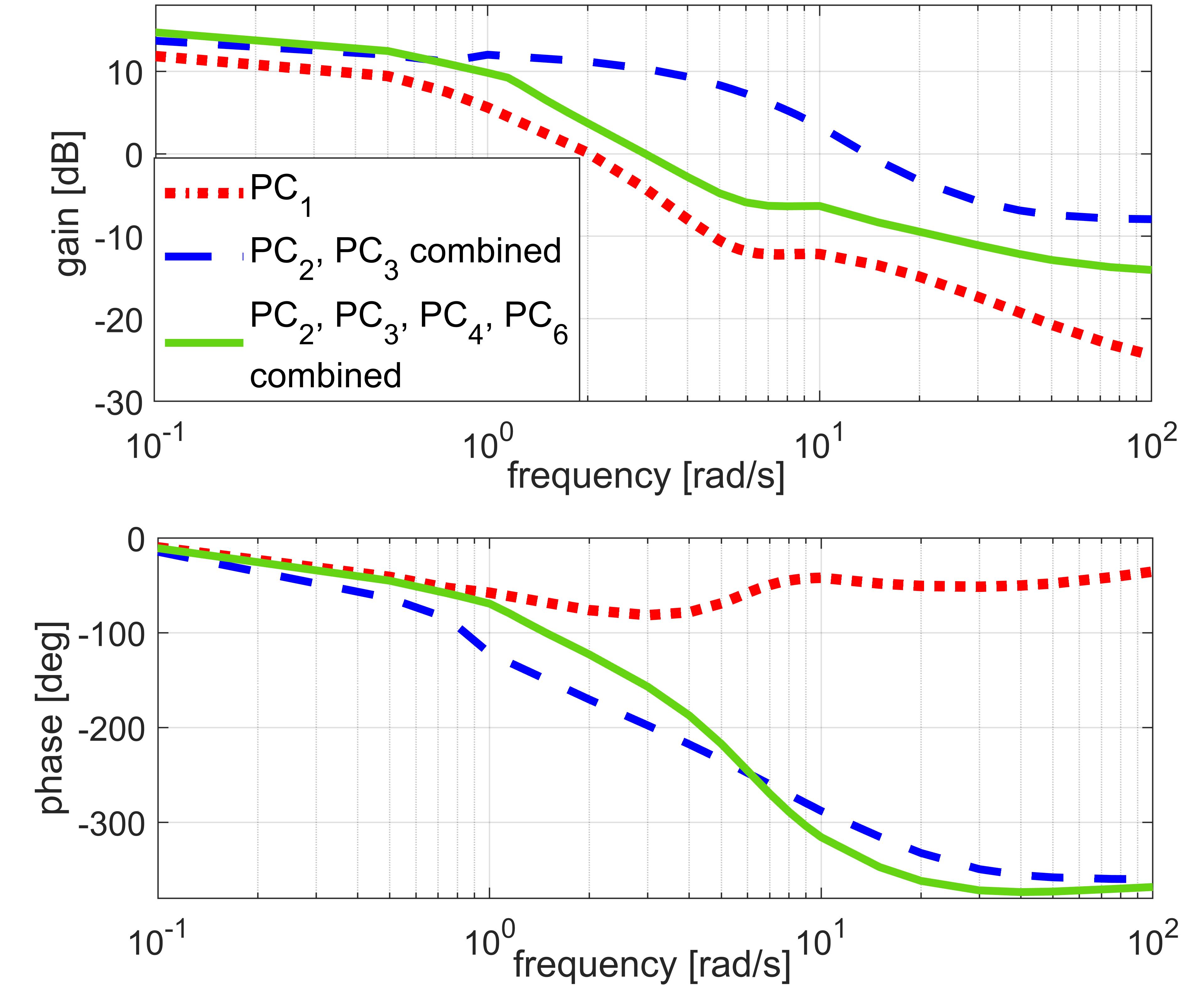}
\caption{Comparison of the pattern angle to yaw rate transfer functions, generated by different combinations of Principal Components extracted from the rotations experiment with the Elite Skydiver. A combination of $PC_2$ and $PC_3$ is defined in Eq. \eqref{eq11}. A combination of $PC_2$, $PC_3$, $PC_4$, $PC_6$ is defined according to Eq. \eqref{eq18}
}
\label{fig15}
\end{figure}

Notice that the derivative of the reference signal in Eq. \eqref{eq16} is known a-priori, as it is related to the intended manoeuvre. The closed loop on the derivative of the yaw rate error, however, may not be possible for human implementation. It requires to sense/ compute the derivative of the actual yaw rate and filter out the noise in real time, what is most likely beyond  human abilities. For this reason, we used a feed-forward of the reference signal derivative combined with a proportional controller, instead of a conventional (Proportional-Derivative) PD control. 

Interestingly, there is another way to combine different movement components. After exploring other Principal Components, we noticed that $PC_2$ in combination with $PC_4$ for right turns and $PC_3$ in combination with $PC_6$ for left turns allows to construct a set of movement patterns, defined by the parameter $k$:
\begin{equation}
\begin{gathered}
    pose(t)= N_{pose}+ \\
\begin{cases}
    \lvert \alpha(t)\rvert\cdot (PC_2-k \cdot PC_4), & \text{if }  \alpha(t)<0\\
    \alpha(t)\cdot (PC_3-k \cdot PC_6),              & \text{otherwise}
\end{cases}
\end{gathered}
\label{eq18}
\end{equation}
where $\alpha(t)$ is the input signal and $k$ is the factor reflecting to what extent the additional patterns $PC_4$ and $PC_6$ are involved.

Movement patterns in this set produce plants that can acquire any behaviour in between the two extremes: the agility of $PC_2$, $PC_3$, and the stability of $PC_1$. The frequency response shown in Fig. \ref{fig15} is for the value of $k=0.8$, whereas it is possible to obtain any response in between the most agile and the most stable behaviour, thus tuning the agility-stability trade-off in a continuous way! This strategy to combine movement components we termed the \textit{Synergy}.

Recall the synergy of the Instructor's Principal Components $PC_1$, $PC_4$ given in Eq. \eqref{eq10} for prevention of orbiting (see Fig. \ref{fig5}): A similar synergy can be observed between the movement components of the Elite Skydiver, see Online Resource \ref{res14}. Moreover, synergies of movement components were identified from analysis of other manoeuvres, for example, the side slides performed by the Elite Skydiver. The details are given in Online Resource \ref{res17}, along with the PCA of additional manoeuvres performed by the Instructor, Student, and Elite Skydiver.

\section{Discussion}\label{sec5}

\subsection{Practical perspective}\label{sec51}

The key idea of our technique analysis method is testing the extracted PCA movement components of the trainees in a Skydiving Simulator and generating transfer functions from a pattern angle to a physical variable associated with a given pattern. 

Simulation tests show what manoeuvre is generated by each PCA movement component, which joint rotation DOFs have a dominant influence, and how many PCA movement components were required to perform the given task or exercise. This information reflects the skill level of the trainee and the major pitfalls in his technique, as was demonstrated in our study case. We suggest to interpret the analysis of the Student's turning patterns as the evolution of his movement repertoire: the old habits ($PC_1$, $PC_3$), exploration ($PC_8$), and on-going progress ($PC_6$).  Movement components that represent exploration by the trainee of the interaction between his body and the airflow usually include a particularly strong movement (a large value in vector $PC_i$ relative to others) of a certain limb, and in the simulation usually produce a turn in only one direction. The progress is represented by movement components that include more unlocked DOFs relative to the most dominant PCA movement component, and in the simulation produce faster turns with less undesirable horizontal motion. We believe that one of the most important tasks of a skydiving coach would be identifying an emergence of such components and triggering the student’s body to utilize them more during next training sessions. The trigger can be found by analysing the PCA control signals of other movement components and the focus of attention at the moment when this component was first activated. For example, in our case, $PC_6$ was triggered when prior to starting the turn the skydiver came to a full stop, increased fall-rate, and kept a visual reference to an object outside of the tunnel at his initial heading. Thus, in order to potentially accelerate learning aerial rotation, the coach could give this student cues to stabilize his initial heading in front of the coach, arch his back, and keep eye contact with the coach. The coach can give a sign to start turning once these conditions are fulfilled. Next, the coach can expect that the analysis of the following training sessions will show that the preferable movement component is more dominant. 

The second potentially useful tool for coaches is constructing transfer functions reflecting the dynamic properties of the body in free-fall actuated by a PCA movement component under investigation. As we have seen in the above analysis the Bode plot of such a transfer function shows what dynamic properties need improvement in order to facilitate for the trainee the implementation of a given manoeuvre. Moreover, it is possible to compare the instructor's and the student's transfer functions, and to check what modifications of the student's movement pattern can reduce the gap between them. Notice, that in skydiving, comparing the movement patterns directly may be less useful, as well as mimicking the movement patterns of a skilled model like the instructor. The reason is that the aerodynamic forces and moments generated by moving a certain limb greatly depend on the individual body shape, height, weight, the type of jumpsuit, and the neutral pose, which in its turn depends on the body flexibility and centre-of-gravity location. Thus, the same movement of a certain student's and instructor's limb can cause a different type of motion in a 3D space, or the same type of motion but in the opposite direction. In this way, from the student's PCA alone the coach cannot know what tips to give him, as there is no 'template' or a 'correct way' to perform a move. Whereas, the tool proposed above allows the coach to know what performance change will likely be caused by a change in movement pattern. In our study case, the coach can recommend the Student to pay attention that his torso tilts rather than rotates, to allow a larger range of movement in the elbow joint, and, finally, to pay attention that in the neutral pose the forearms are in the same plane as the torso. These changes will enable the Student to be less sensitive to the turbulence of the airflow, and generate faster turns while not losing stability. The simulated improvements need to be experimentally verified and compared to other practice and instruction conditions.

 Introducing the desired changes into practice should be gradual (one change at a time), and, while performing the manoeuvre, the focus of attention should still be external, e.g. looking at the instructor as mentioned above. When initiating a movement (e.g. a rotation) the trainee should consciously apply the desired change (e.g. torso tilt) and pay attention to how it feels and how the airflow around the body is redistributed. This new initial response, according to fundamental somatic practices (\cite{sport_brodie2012dance}), will help to break previously acquired ‘bad habits’ and trigger emergence of new movement patterns. 
 
\subsection{Theoretical perspective} \label{sec52}

The main hypothesis of this research, demonstrated in the study case, explains the learning process of a new motor skill from the point of view of control theory. 
The learning begins with a primitive body actuation leading to a problematic plant, and forcing novices to act in an open loop since the required controller would have a dynamic complexity that is  hard to realize by the human motor mechanisms.  Over time, the learning converges to operating in a closed loop via a simple control law since an improved body actuation  provides the plant with the desirable dynamic properties. When the skill level is sufficiently high, the stability-agility trade-off becomes shifted towards the latter in order to enable performing advanced manoeuvres. The plant, inherently unstable in this case, is controlled in open loop, while the feed-forward gain is learnt by trial-and-error. This process is very time-consuming: athletes repeat each competition move thousands of times in the wind tunnel.  
Probably, for this reason the specialization in skydiving is so narrow. There are many disciplines, and at each discipline skydivers compete only at a very limited number of specific manoeuvres. For example, the oldest skydiving discipline called 'Style and Accuracy' included only two manoeuvres: turns and back loops.

We have discovered that for most manoeuvres a significant improvement of plant dynamics can be achieved by the means of modifying the engagement of only a few (or even just one) DOFs. In Sect. \ref{sec43} it was the torso tilt, which we termed the Key DOF. We observed that the key DOF is usually proximal: one of the torso DOFs, shoulders, and hips. This can be expected since a small change in these joints creates a significant displacement of the distal limbs (due to a large lever arm), allowing to place them in a more efficient position relative to the airflow. This assumes, however, that the distal DOFs are unlocked, i.e. engaged in movement.


Thus, the insight into the Bernstein's DOF problem, emerged from our research, is the following: Movement patterns required for a skilful task execution incorporate a large amount of  body DOFs in order to account for many aspects of the environmental dynamics. Human kinematic redundancy is utilized for achieving a dynamic system with good handling characteristics: sufficient bandwidth and stability margins, high frequency disturbance attenuation, no cross-coupling, zero steady-state error for step inputs, fast rise time, small overshoot, and other qualities depending on the desired manoeuvre or task. 

Simulations show that it is impossible to produce a plant with good handling qualities if the movement pattern that actuates the body engages only 1-2 joints rotation DOFs, even if these DOFs are sufficient for initiating a desired manoeuvre, e.g. a turn. This is the reason why the current skydiving training is so hard and protracted. The movements that are taught to students are simple, but it’s very hard to use them in reality, while the efficient movements that are convenient for manoeuvring are not taught, as it is difficult to explain them. In order to develop efficient movement patterns the body needs to learn the interaction with the environment: the aerodynamics of free-fall. This is achieved through performing free-fall manoeuvres, what is extremely difficult for novices, since they are trying, according to our model, to control a plant with very poor dynamic characteristics. This results in stability loss and spending most of the training session (free-fall time) in attempts to regain it. This vicious circle can be broken, theoretically, in two ways, which are the primary directions for our future work: 

The first option is showing the trainee manoeuvres that can be performed using the simple movements. Firstly, determine (by the means of simulations) the performance envelope of a trainee given his current movement repertoire. Secondly, design individual exercises that are inside the trainee's performance envelope. The trainee is thus given tasks that we know are achievable, i.e. we prevent him from losing stability and wasting training time. The trainee’s body will acquire a feel for flight dynamics thus triggering the CNS to produce new movement patterns, utilizing additional DOFs, and extending the performance envelope.

The second option is finding a way to teach the more efficient movements. For example, extract PCA movement components from experiments with an experienced skydiver, as we showed in the sections above. Find by the means of the simulator what joints rotation DOFs are most significant for each of the expert’s dominant movement components. Create an animation of a pattern that engages only these significant DOFs for the trainee to view. If the combination of the selected DOFs is not intuitive for execution, viewing the animation will not be sufficient. In this case, it will be advantageous to practice the pattern execution on the ground. The trainee can wear a body motion tracking suit and his movements can be continuously displayed in the same environment as the pattern’s animation. The trainee can practice until his body posture continuously coincides with the posture animating the desired pattern. After that, the training can proceed, as described in the former paragraph. In our future work, this approach will be experimentally verified and compared to other practice and instruction conditions.

The method presented in this work can be applied to other activities and sports disciplines. The modelling part has to be specific for the activity under investigation: a set of equations propagating the athlete’s motion in the 3D world (velocity, position, orientation, i.e. state variables) as a function of his instantaneous body coordination (relevant DOFs).  Once an appropriate model of the activity dynamics is formulated, the tools analysing movement components and predicting their potential improvements apply. The transfer function construction procedure is general for all applications, only the input and output variables are problem-specific. Complex manoeuvres might be defined by more than one state variable, and might involve more than one movement component. In this case several transfer functions should be constructed: for each possible pair of movement component – state variable, as e.g. in \cite{article}.     

\section{Conclusion}\label{sec6}

The presented method of movement patterns analysis provided some useful insights into the natural mechanisms of utilizing motor equivalence for performing skydiving manoeuvres. Three key mechanisms were identified:
\begin{enumerate}
    \item \textbf{Plant shaping via DOFs synergies}: mechanism of constructing movement patterns appropriate for specific manoeuvres.
    \item \textbf{Superposition of movement patterns}: utilising several movement patterns during a specific manoeuvre, while each pattern is engaged via a different control strategy.
    \item \textbf{Synergy of movement patterns}: utilising a specific combination of different movement patterns, which is aimed to achieve a desired trade-off between various dynamic characteristics required for a given task.
\end{enumerate}
The most important discovery is concerned with an impact of multiple body DOFs on the dynamics of body flight. Conventional motor control computational models \citep{berret2011evidence, motor_haith2013theoretical, motor_todorov2005task, motor_wolpert2011principles} treat the body DOFs as redundant actuators. This concept inherently assumes that the system dynamic behaviour is the same for all possible distributions of the total control effort between those actuators. However, we have seen in both simulations and experiments that the dynamics of body flight highly depends on the choice of DOFs involved in movement. We propose, therefore, to view the classical Berstein's DOFs problem not only as a \textit{kinematic redundancy}, but also from a dynamics perspective. Our hypothesis is the following:

\vspace{0.3cm}

\textit{The multiple body DOFs are not necessarily redundant, but they are needed for shaping the plant dynamics to enable performing desired manoeuvres with a simple control law.}

\vspace{0.3cm}

This way, every movement pattern in the athlete's movement repertoire is dedicated for a specific manoeuvre and provides the plant dynamics, suitable or maybe even optimal for this manoeuvre. This means that an automatic pilot in the diagram shown in Fig. \ref{fig2} has to be designed specifically for each movement pattern, which is chosen for a manoeuvre or task. This is the key idea of our multiple actuator control approach for mathematical modelling of motor control, whereby each movement pattern, or a specific Synergy of movement patterns, is treated as a separate actuator.

\backmatter

\bmhead{Supplementary information}

The following supplementary files are provided:
\begin{enumerate}
\item Video recording of the wind tunnel rotations experiment with the Instructor \label{res1}
\item Video recording of the wind tunnel rotations experiment with the Student \label{res2}
\item Video recording of the wind tunnel rotations experiment with the Elite Skydiver \label{res3}
    \item Animation of the Instructor's first Principal Component, extracted by the means of PCA \label{res4}
    \item Plot of the control signals of the first six Principal Components extracted from the experiment with the Student \label{res5}
    \item Plot of the yaw rate reconstructed by the skydiving simulator using different amounts of Principal Components extracted from the experiment with the Student \label{res6}
    \item Animation of the Student's first Principal Component, extracted by the means of PCA \label{res7}
    \item Animation of the Student's third Principal Component, extracted by the means of PCA \label{res8}
    \item Animation of the Student's sixth Principal Component, extracted by the means of PCA \label{res9}
    \item Animation of the Student's eighth Principal Component, extracted by the means of PCA \label{res10}
    \item Animation of the first Principal Component of the Elite Skydiver, extracted by the means of PCA \label{res11}
     \item Animation of the second Principal Component of the Elite Skydiver, extracted by the means of PCA \label{res12}
      \item Animation of the third Principal Component of the Elite Skydiver, extracted by the means of PCA \label{res13}
      \item Plots of the PCA results of the experiment with the Elite Skydiver, and yaw rate reconstruction in simulation using the obtained Principal Components \label{res14}
      \item Recording of the graphical output of the open-loop simulation driven by the first Principal Component of the Elite Skydiver \label{res15}
      \item Recording of the graphical output of the open-loop simulation driven by the combination of the second and third Principal Components of the Elite Skydiver \label{res16}
      \item  PCA of additional manoeuvres performed by the Instructor, Student, and Elite Skydiver \label{res17}
\end{enumerate}

\section*{Declarations}

\bmhead{Conflict of interest}
The authors have no relevant financial or non-financial interests to disclose.

\bmhead{Funding}
No funds, grants, or other support was received.

\bmhead{Author contributions}
The manuscript draft preparation, and data collection and analysis were performed by Anna Clarke. The study supervision and the manuscript editing was performed by Per-Olof Gutman. Both authors read and approved the final manuscript.

\bmhead{Ethics approval}
The experiments were approved by the Technion Institutional Review Board and Human Subjects Protection.

\bmhead{Consent to participate}
The participants were informed of the aims and procedures of the experiments, and signed an informed consent form.

\bmhead{Consent for publication}
The participants have agreed to publication of the motion data collected during the experiments, the data processing results, and videos recorded during the experiments.

\bmhead{Availability of data and materials}
All data (Xsens motion data collected during the experiments) is available from 4TU.Centre  for  Research  Data:  
          \url{https://doi.org/10.4121/uuid:c56e609a-e6b1-463c-8a61-4386d8c8dbb0}

\bmhead{Code availability}
The code is available from Figshare Collection: \textit{Matlab Code for Skydiving Technique Analysis.}  \url{https://doi.org/10.6084/m9.figshare.c.5692768.v1}




\end{document}